\documentclass[a4paper,fleqn,usenatbib]{mnras}

\usepackage{latexsym}
\usepackage{amssymb}
\usepackage{amsfonts}
\usepackage{amsmath}
\usepackage{bm}
\usepackage{graphicx}
\usepackage{subfigure}
\usepackage{times}
\usepackage{units}
\usepackage{hyperref}
\usepackage{multirow}
\usepackage{comment}
\usepackage{ulem}
\usepackage{hyperref}
\usepackage{graphicx}

\usepackage{pifont} 

\usepackage{enumitem}
\setlist[itemize]{noitemsep, topsep=0pt}

\newcommand{\be}{\begin{equation}}
\newcommand{\ee}{\end{equation}}
\newcommand{\bea}{\begin{eqnarray}}
\newcommand{\eea}{\end{eqnarray}}
\newcommand{\bel}{\begin{align}}
\newcommand{\eel}{\end{align}}

\renewcommand{\cite}[1]{\citep{#1}}
\def\igscm{{\rm cm^{2}\,g^{-1}}}
\def\ergsg{{\rm erg\,g^{-1}\,s^{-1}}}
\def\Msun{{\rm M_{\odot}}}
\def\GMc2{{\rm G M_{\odot} c^{-2}}}
\def\Rph{R_{\rm ph}}

\def\eps{\epsilon}

\def\B{\mathcal{B}}

\def\Mo{{\rm M_{\odot}}}
\def\Me{{m}}
\def\Metot{{M_{\rm ej}}}

\def\Tfni{{ T^{\rm Ni}_{\rm floor}}}
\def\Tfla{{ T^{\rm LA}_{\rm floor}}}
\def\bdphi{{\boldsymbol{\theta}}}

\def\vrms{{v_{\rm rms}}}
\def\klow{\kappa_{\rm low}}
\def\khigh{\kappa_{\rm high}}
\def\lt{{\tilde\Lambda}}
\def\rof{{R_{1.4}}}
\def\mg{{\rm mag}}

\def\kt2{\kappa^\text{T}_2}

\usepackage{pifont} 
\usepackage{color}
\definecolor{cyan}{rgb}{0,0.9,0.9}
\definecolor{orange}{rgb}{0.9,0.5,0}
\definecolor{magenta}{rgb}{1,0,1}
\definecolor{purple}{rgb}{0.8,0.4,0.8}
\definecolor{gray}{rgb}{0.8242,0.8242,0.8242}

\newcommand{\oldtxt}[1]{\sout{{{#1}}}}
 
\title[AT2017gfo: model selection 
of multi-component kNe]{AT2017gfo: Bayesian inference 
	and model selection 
  of multi-component kilonovae
   and constraints on the neutron star equation of state}

\author[M. Breschi et al.]{{Matteo {Breschi}$^{1}$},
{Albino {Perego}$^{2,3}$},
{Sebastiano {Bernuzzi}$^{1}$},
{Walter {Del Pozzo}$^{4,5}$},
\newauthor
{Vsevolod {Nedora}$^{1}$},
{David {Radice}$^{6,7,8}$},
{Diego {Vescovi}$^{9,10,11}$}
\\
{${}^1$Theoretisch-Physikalisches Institut, Friedrich-Schiller-Universit{\"a}t Jena, Fr\"obelstieg 1, 07743, Jena, Germany}\\
{${}^2$Dipartimento di Fisica, Universit\'a di Trento, Via Sommarive 14, 38123, Trento, Italy}\\
{${}^3$INFN-TIFPA, Trento Institute for Fundamental Physics and Applications, via Sommarive 14, 38123, Trento, Italy}\\
{${}^4$Dipartimento di Fisica ``Enrico Fermi'', Universit\'a di Pisa, Largo B. Pontecorvo 14, 56127, Pisa, Italy}\\
{${}^5$INFN, Sezione di Pisa, Largo B. Pontecorvo 14, 56127, Pisa, Italy}\\
{${}^6$Institute for Gravitation \& the Cosmos, The Pennsylvania State University, University Park, PA 16802, USA}\\
{${}^7$Department of Physics, The Pennsylvania State University, University Park, PA 16802, USA}\\
{${}^8$Department of Astronomy \& Astrophysics, The Pennsylvania State University, University Park, PA 16802, USA}\\
{${}^9$Gran Sasso Science Institute, Viale F. Crispi 7, 67100, L’Aquila, Italy}\\
{${}^{10}$INFN, Sezione di Perugia, Via A. Pascoli 23, 06123, Perugia, Italy}\\
{${}^{11}$INAF, Observatory of Abruzzo, Via M. Maggini, 64100, Teramo, Italy}}

\date{
	Accepted 2021 May 3. Received 2021 May 3; in original form 2021 January 7
}

\begin{document}
\label{firstpage}
\maketitle
	
\date{\today}

\begin{abstract} 
	The joint detection of the gravitational wave GW170817, of the short $\gamma$-ray 
burst GRB170817A and of the kilonova AT2017gfo, 
generated by the the binary neutron star merger
observed on August 17, 2017, is a milestone in multimessenger astronomy
and provides new constraints on the neutron star equation of state.
We perform Bayesian inference and model selection on AT2017gfo using
semi-analytical, multi-components models that also account for
non-spherical ejecta.
Observational data favor anisotropic geometries to
spherically symmetric profiles,
with a log-Bayes' factor of ${\sim}10^{4}$, and
favor multi-component models against single-component ones. 
The best fitting model is an anisotropic three-component composed of
dynamical ejecta plus neutrino and viscous winds.
Using the dynamical ejecta parameters inferred from the best-fitting
model and numerical-relativity relations connecting 
the ejecta properties to the
binary properties, we constrain the binary mass ratio to $q<1.54$ and
the reduced tidal parameter to $120<\tilde\Lambda<1110$. 
Finally, we combine the predictions from AT2017gfo with those from
GW170817, constraining the radius of a neutron star of $1.4~\Mo$ 
to $12.2\pm0.5~{\rm km}$
($1\sigma$ level).
This prediction could be further strengthened by
improving kilonova models with numerical-relativity information.
\end{abstract}



\section{Introduction} 
\label{sec:intro}

On August 17, 2017, the ground-based interferometers of LIGO and Virgo~\cite{Aasi:2013wya,TheLIGOScientific:2014jea,TheVirgo:2014hva}
detected the first gravitational-wave (GW) signal 
coming from a binary neutron star (BNS) merger,
known as GW170817~\cite{TheLIGOScientific:2017qsa}.
GW170817 was followed by a short gamma-ray burst (GRB) GRB170817A~\cite{Monitor:2017mdv,Savchenko:2017ffs},
which reached the space observatories Fermi~\cite{TheFermi-LAT:2015kwa} 
and INTEGRAL~\cite{Winkler_2011}
${\sim}1.7\,{\rm s}$ after coalescence time.
Eleven hours later, several telescopes
 started to collect photometric and spectroscopical data from AT2017gfo,  an unprecedented electromagnetic (EM) kilonova transient
~\cite{Coulter:2017wya,Chornock:2017sdf,Nicholl:2017ahq,Cowperthwaite:2017dyu,Pian:2017gtc,Smartt:2017fuw,Tanvir:2017pws,Tanaka:2017qxj,Valenti:2017ngx}
coming from a coincident region of the sky.
Kilonovae (kNe) are quasi-thermal EM emissions interpreted as distinctive signature of $r$-process nucleosynthesis in the neutron-rich matter ejected from the merger and from the subsequent BNS remnant evolution~\cite{Smartt:2017fuw,Kasen:2017sxr,Rosswog:2017sdn,Metzger:2019zeh,Kawaguchi:2019nju}.
The follow up of the source lasted for more than a month
and included also non-thermal emission from the GRB170817A afterglow~\citep[e.g.,][]{Nynka:2018vup,Hajela:2019mjy}.

The combined observation of GW170817, GRB170817A and AT2017gfo decreed
the dawn of multimessenger astronomy with compact binaries~\cite{GBM:2017lvd}.
From these multimessenger observations it is possible to infer unique 
information on the unknown equation of state (EOS) of neutron star (NS) matter, 
\citep[e.g.][]{Radice:2016rys,Margalit:2017dij,Bauswein:2017vtn,Radice:2017lry,Dietrich:2018upm}.
Indeed, the EOS determines the tidal polarizability
parameters that describe tidal interactions during the
inspiral-merger and characterize the GW signal~\cite{Damour:2012yf,Bernuzzi:2014kca}.
It also determines the outcome of BNS mergers~\citep[e.g.][]{Shibata:2005ss,Bernuzzi:2014owa,Bernuzzi:2020txg}
and the subsequent postmerger GW signal from the remnant~\citep[e.g.][]{Bauswein:2014qla,Bernuzzi:2015rla,Zappa:2019ntl,Agathos:2019sah,Breschi:2019srl}.
At the same time,
the amount of mass, the velocity, and the composition of the ejecta
are also strongly dependent on the EOS, that has an imprint on
the kN signature,
e.g.~\cite{Hotokezaka:2012ze,Bauswein:2013yna,Radice:2018pdn,Radice:2018xqa}.

The spectrum of AT2017gfo was recorded from ultraviolet (UV) to near infrared (NIR) 
frequencies~\citep[e.g.,][]{Pian:2017gtc,Nakar:2018cbe},
and the observations showed several characteristic features. 
At early stages, the kN was very bright and its spectrum peaked in the blue band 1 day after 
the merger (blue kN).
After that, the peak of the spectrum moved towards larger wavelengths, peaking at NIR frequencies 
between five to seven days after merger (red kN).
Minimal models that can explain these features require 
more than one component. In particular,
minimal fitting models assume spherical symmetry and include
a lathanide-rich ejecta responsible for the red kN, 
typically interpreted as dynamical ejecta, and another ejecta with material partially reprocessed by weak
interaction, responsible for the blue component~\citep[e.g.,][]{Villar:2017wcc}.
Numerical relativity (NR) simulations show that the geometry profiles of the 
ejecta are not always spherically symmetric and their distributions
are not homogeneous~\cite{Perego:2017wtu}.
Moreover, NR simulations also indicate the presence of multiple
ejecta components, from the dynamical to the disk winds ejecta~\cite{Rosswog:2013kqa,Fernandez:2014bra,Metzger:2014ila,Perego:2014fma,Nedora:2019jhl}.
Therefore, this information has to be taken into account during the
inference of the kN properties.
  
The modeling of kNe is a challenging problem,
due to the complexity of the underlying physics, which
is affected by a diverse interactions and scales~\citep[see][ and references therein]{Metzger:2019zeh}.
Together with the choice of ejecta profiles,
the lack of a reliable description of the radiation transport 
is a relevant source of uncertainties in the modeling of kNe, due to the 
incomplete knowledge on the thermalization processes~\cite{Korobkin:2012uy,Barnes:2016umi}
and on the energy-dependent photon opacities in $r$-process matter~\cite{Tanaka:2019iqp, Even:2019nbj}.
Current kN models often use either simplistic ejecta profiles or simplistic radiation 
schemes, ~\citep[e.g.,][]{Grossman:2013lqa,Villar:2017wcc,Coughlin:2017ydf,Perego:2017wtu}.
Given the challenges and uncertainties associated to the theoretical prediction of kN features, 
Bayesian inference and model selection of the observational data can provide important insights 
on physical processes hidden in the kN signature.

In this work, we explore model selection in geometrical and
ejecta properties using simplified light curve (LC) models, that nonetheless capture the key features of the problem. 
The inference results are then employed to derive constraints on the neutron star EOS.
In Sec.~\ref{sec:knmodel}, we describe the semi-analytical model and the ejecta components used in our analysis. 
In Sec.~\ref{sec:method}, we recall the Bayesian framework for model selection, highlighting the
choices of the relevant statistical quantities, such as likelihood function and prior distributions.
In Sec.~\ref{sec:results}, we discuss the inference on AT2017gfo,
critically examining the posterior samples in light of
targeted NR simulations~\cite{Perego:2019adq,Nedora:2019jhl,Endrizzi:2019trv,Nedora:2020pak,Bernuzzi:2020txg} and previous analyses. 
In Sec.~\ref{sec:eos}, we discuss new constraints on the NS EOS focusing first on mass ratio and reduced tidal parameter for the source of GW170817, and then on the neutron star radius $\rof$.
We conclude in Sec.~\ref{sec:conclusions}.


\section{Kilonova model} 
\label{sec:knmodel}

In this section, we first summarize basic analytical results and scaling relations that characterize the kN emission, and then describe in detail the models we employ for the ejecta components and LC calculations.

\subsection{Basic features}
\label{subsec:lc}

Let us consider a shell of ejected matter
characterized by a mass density $\varrho$, 
with total mass $\Me$
and gray opacity $\kappa$ (mean cross section per unit mass). The shell is in homologous expansion symmetrically with respect to the equatorial plane at velocity $v$, such that its mean 
radius is $R \sim v t $ after a time $t$ following the merger. 
Matter opacity to EM radiation can be expressed in terms of the optical depth, $\tau$, which is estimated as $\tau \simeq \varrho \kappa R$.
After the BNS collision, when matter becomes unbound and $r$-process nucleosynthesis occurs, the ejecta are extremely hot, $T\sim10^9~{\rm K}$ \citep[e.g.][]{Mendoza-Temis:2014mja,Wu:2016pnw,Perego:2019adq}.
However, at early times the thermal energy is not dissipated efficiently since the environment is optically thick ($\tau \gg 1$) and photons diffuse out only on the diffusion timescale until they reach the photosphere ($\tau = 2/3$).
As the outflow expands, its density drops ($\rho \propto t^{-3}$) and the optical depth decreases.

The key concept behind kNe is that photons can contribute to the EM emission at a given time $t$ if they diffuse on a timescale comparable to the expansion timescales, i.e., if they escape from the shells outside $R_{\rm diff}$, where $R_{\rm diff}$ is the radius at which the diffusion time $t_{\rm diff} \simeq R\tau /c $ equals the dynamical time $t$ ~\cite{Piran:2012wd,Grossman:2013lqa,Metzger:2019zeh} .
In the previous expression, $c$ is the speed of light.
Since $t_{\rm diff} \propto t^{-1} $, a larger and larger portion of the ejecta becomes transparent with time.
The luminosity peak of the kN
occurs when the bulk of matter 
that composes the shell becomes transparent.
As first approximation, the characteristic timescale at which the light 
curve peaks is commonly estimated~\cite{Arnett:1982} as:
\begin{equation}
\label{eq:tpeak}
t_{\rm peak} = \sqrt{\frac{3\Me \kappa}{4\pi \beta v c}}
\,,
\end{equation}
where the dimensionless factor $\beta$ depends on the density profile of the ejecta.
For a spherical symmetric, homologously expanding ejecta ($\beta \simeq 3 $) with mass $\Me = 10^{-2}~\Mo $, 
velocity $v = 0.1~c$ and opacity in the range $\kappa \simeq 1{-}50~\igscm$,
which are typical values respectively for lanthanide-free and for lanthanide-rich matter~\cite{Roberts:2011xz,Kasen:2013xka},
Eq.~\eqref{eq:tpeak} predicts a characteristic $t_{\rm peak}$ in the range $1$--$10~{\rm days}$~\cite{Abbott:2017wuw}.

In the absence of a heat source, matter would simply cool down through adiabatic expansion.
However, the ejected material is continuously heated by the radioactive decays of the $r$-process yields, which provide
a time dependent heating rate of nuclear origin. 
An additional time dependence is introduced by the thermalization efficiency, i.e. the efficiency at which this nuclear energy, released in the form of supra-thermal particles (electrons, daughter nuclei, photons and neutrinos), thermalizes within the expanding ejecta \citep[see, e.g.,][]{Metzger:2011bv,Korobkin:2012uy,Barnes:2016umi,Hotokezaka:2018aui}.

\subsection{Light Curves}
\label{subsec:lc}

The kN LCs in our work are computed using the multicomponent, anisotropic semi-analytical {\tt MKN} model first introduced in Ref.~\cite{Perego:2017wtu} and largely based on the kN models presented in Refs.~\cite{Grossman:2013lqa} and \cite{Martin:2015hxa}
(see also \citet{Barbieri:2019sjc}).
The ejecta are either spherical or axisymmetric with respect to the 
rotational axis of the remnant, and symmetric with respect to the equatorial plane.
The viewing angle $\iota$ is measured as the angle between the 
rotational axis and the line of sight of the observer.

For each component the ejected material is described through 
the angular distribution of its ejected mass, $\Me$, root-mean-square (rms) radial velocity, $\vrms$, and opacity, $\kappa$.
In axisymmetric models, the latter quantities are functions of the polar angle $\theta$, measured from the rotational axis and discretized in $N_\theta=30$ angular bins evenly spaced in $\cos{\theta}$.
Additionally, within each ray, matter is radially distributed with a stationary profile in velocity space, $\xi(v)$ such that $\xi(v) \propto (1 - \left(v/v_{\rm max}\right)^{2})^{3}$, where $\xi(v) {\rm d}v$ is the matter contained in an infinitesimal layer of speed $\left[v,v+{\rm d}v\right]$, and $v_{\rm max}=v_{\rm max}(\vrms)$ is the maximum velocity at the outermost edge of the component.
The characteristic quantities $\varrho$, $v$ and $\kappa$ are then evaluated for every bin according to the assumed input profiles.
For every bin, we estimate the emitted luminosity using the radial model described in Ref.~\cite{Perego:2017wtu} and in \S{4} of Ref.~\cite{Barbieri:2019kli} (see also \cite{Barbieri:2019sjc}).
In particular, the model assumes that the luminosity is emitted as thermal radiation from the photosphere (of radial coordinate $\Rph$), and the luminosity and the photospheric surface determine the effective emission temperature, $T_{\rm eff}$ through the Stefan-Boltzmann law.
We expect this assumption to be well verified at early times (with a few days after merger), while deviations from it are expected to become
more and more relevant for increasing time.

The time-dependent nuclear heating
rate $\eps_{\rm nuc }$ entering these calculations is approximated by an analytic fitting formula, derived from detailed nucleosynthesis calculations~\cite{Korobkin:2012uy},
\be
\label{eq:epsnuc}
\eps_{\rm nuc}(t)= \eps_0 \, \frac{\eps_{\rm th}(t)}{0.5} \, \eps_{\rm nr}(t) \,\left[ \frac{1}{2} - \frac{1}{\pi} \arctan\left(\frac{t-t_0}{\sigma}\right)\right]^{\alpha}\,,
\ee
where $\sigma = 0.11~{\rm s}$, $t_0 = 1.3~{\rm s}$, $\alpha=1.3$ and $\eps_{\rm th}(t)$ is the thermalization
efficiency tabulated according to Ref.~\cite{Barnes:2016umi}. The heating factor $\eps_{\rm nr}(t) $ is introduced as in Ref.~\cite{Perego:2017wtu} to roughly improve the behavior of Eq.~\eqref{eq:epsnuc} in the regime of mildly neutron-rich matter (characterized by an initial electron fraction $Y_e \gtrsim 0.25$), \citep[see, e.g.][]{Martin:2015hxa}:
\be
\label{eq:epsnr}
\eps_{\rm nr}(t,\kappa) = \left[1-w(\kappa)\right] + w(\kappa)\,\eps_{Y_e}(t)\,,  
\ee
where $w(\kappa)$ is a logarithmic smooth clump function such that $w(\kappa< 1~\igscm) = 1$ and 
$w(\kappa> 10~\igscm)=0$ and the factor $\eps_{Y_e}(t)$ encodes the dependence on  $Y_e$:
if $Y_e < 0.25$, then $\eps_{Y_e}(t)=1$, otherwise, when $Y_e \ge 0.25$,
\be
\label{eq:epsye}
\eps_{Y_e}(t) =\eps_{\rm min}+{\eps_{\rm max}}{\left[1+ e ^{4(t/t_\eps-1)}\right]}^{-1}\,,
\ee
where $t_\eps = 1~{\rm day}$, $\eps_{\rm min}=0.5$ and $\eps_{\rm max} = 2.5$.

Furthermore, in order to improve the description in the high-frequency bands (i.e., $V$, $U$, $B$ and $g$) within the timescale of the kilonova emission, and following Ref.~\cite{Villar:2017oya}, we introduce a floor temperature, i.e. a minimum value for $T_{\rm eff}$. This is physically related to the drop in opacity due to the full recombination of the free electrons occurring when for the matter temperature drops below $T_{\rm floor}$~\cite{Kasen:2017sxr,Kasen:2018drm}. Under these assumptions, the condition $T_{\rm eff} = T_{\rm floor}$ becomes a good tracker for the photosphere location.
Since kNe are powered by the radioactive decay of 
 different blends of atomic species,
we introduce in our model two floor temperatures, $\Tfni$ and $\Tfla$, that characterize respectively the recombination temperature of lanthanides-free and of lanthanide-rich ejecta.

Eventually, the emissions coming from the different rays are combined to obtain the spectral flux at the observer location:
\be
\label{eq:spectral_flux}
F_{\nu}(\mathbf{n},t) = \int_{\mathbf{n}_{\Omega} \cdot \mathbf{n}> 0} \left( \frac{\Rph(\Omega,t)}{D_L} \right)^2  B_{\nu}(T_{\rm eff}(\Omega,t))~\mathbf{n} \cdot {\rm d}\boldsymbol{\Omega} 
\ee
where $\mathbf{n}$ is the unitary vector along the line of sight, $\mathbf{n}_{\Omega}$ is the unitary vector spanning the solid angle $\Omega$, $D_L$ is the luminosity distance, $\Rph$ is the local radial coordinate of the photospheric surface, and $B_{\nu}(T_{\rm eff})$ is 
the spectral radiance at frequency $\nu$ for a surface of temperature $T_{\rm eff}$. 
Lastly, from Eq.~\eqref{eq:spectral_flux}, it is possible to compute the apparent AB magnitude $\mg_b$ in a given photometric band $b$ as:
\be
\label{eq:mag}
\mg_b(\mathbf{n},t) = -2.5 \log_{10}\left( F_{\nu_b}(\mathbf{n},t) \right)-48.6\,,
\ee
where $\nu_b$ is the effective central frequency of band $b$.

\subsection{Multi-Component Model}
\label{subsec:multicomponents}

\begin{figure}
	\centering 
	\includegraphics[width=0.48\textwidth]{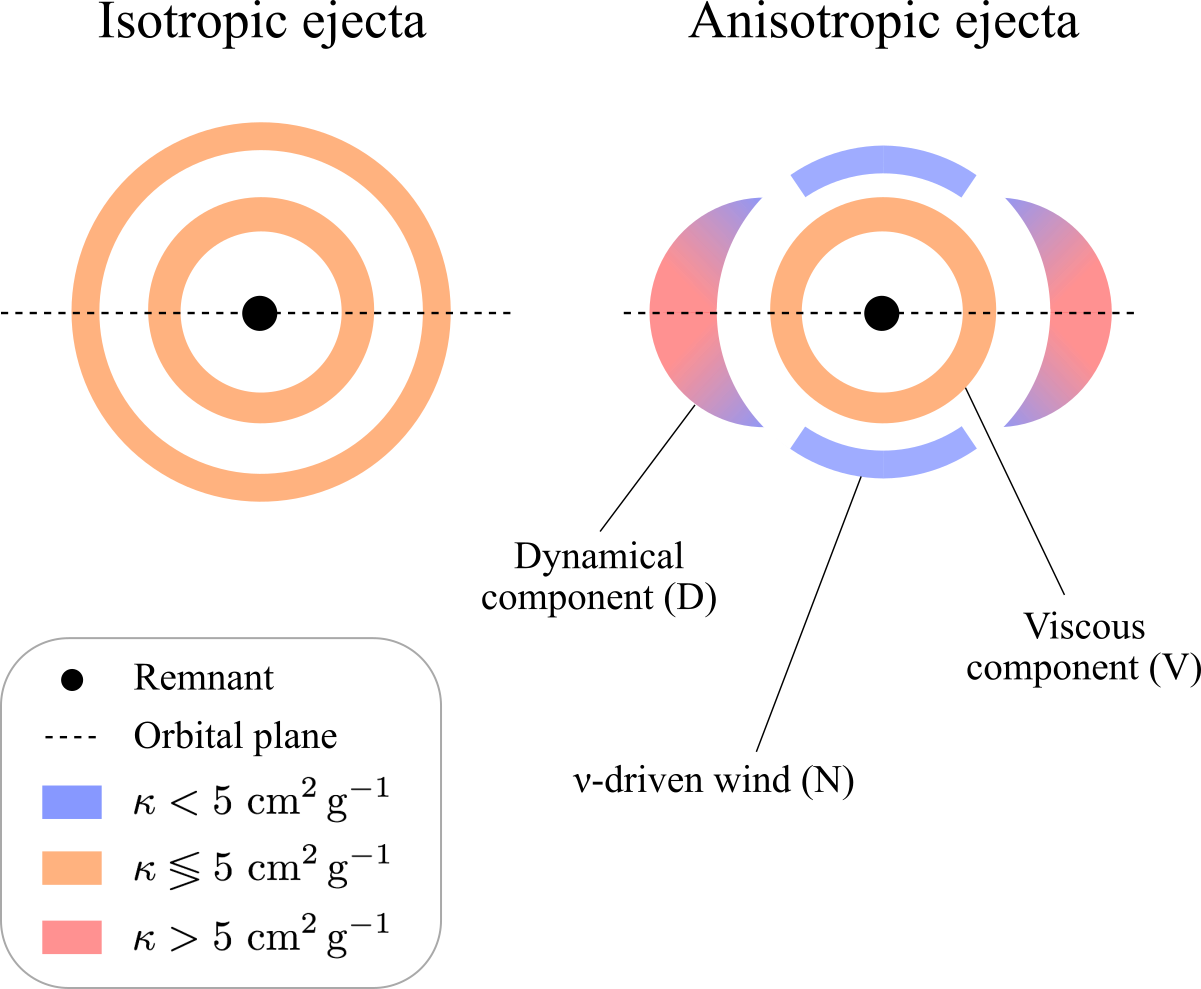}
	\caption{Graphic representation of the analyzed
					ejecta profiles for isotropic and anisotropic cases
					from an azimuthal perspective and for a fixed moment of time.
				   The black dot represents the remnant and the dashed line is the projected orbital
			   		plane of the binary. The shadowed areas describe the ejecta profiles: the shape
		   		    characterizes the mass distribution, while the colors refer to 
		   		    the prior assumptions on the opacity parameter.
		   		    In particular, blue regions denote opacities lower than $5~\igscm$,
		   		    red regions refer to opacities greater than $5~\igscm$,
		   		    and oranges areas indicate a broadly distributed opacity.
	   		    	All shells are isotropically expanding with a constant velocity.}
	\label{fig:cartoon}
\end{figure}

In order to describe the different properties of AT2017gfo
it is necessary to appeal to a multi-component structure for the ejecta producing the kN.
Different components are characterized by different sets of intrinsic parameters, $\Me$, $\vrms$ and $\kappa$, and by their angular distributions with respect to $\theta$.

Given the angular profiles of the characteristic parameters, the physical luminosity produced by each component inside a ray is computed by using the model outlined in the previous section. 
Then, the total bolometric luminosity of the ray
is given by the sum of the single contributions, i.e. $L(t)= \sum_{k} L^{(k)}(t)$ where $k$ runs over the components. The outermost photosphere is the one that determines the thermal spectrum of the emission.
Once $\Rph$ and $T_{\rm eff}$ have been determined, the spectral flux and the AB magnitudes are computed according to Eqs.~(\ref{eq:spectral_flux}) and~(\ref{eq:mag}).

We perform the analysis using two different assumptions on the profiles of the source.
Initially, we impose completely isotropic profiles
for every parameter of every ejecta component.
These cases are labeled as isotropic, `ISO'.
Subsequently, we introduce angular profiles 
as functions of the polar angle
for the mass and opacity parameters, while we keep $\vrms$ always isotropic.
This second case is labeled as anisotropic, `ANI'.
In parallel, we explore models with a different number of components. We always assume the presence of the dynamical ejecta, while we add to them one or two qualitatively different disk-wind ejecta components.

In the following paragraphs, we describe the physical assumptions
on each component and the choice of the prior distributions (see Tab.~\ref{tab:params}).
Fig.~\ref{fig:cartoon} shows a graphical representation of the employed ejecta components.

\paragraph*{Dynamical ejecta $({\rm D})$.} The BNS collision ejects unbound matter on the dynamical timescale, whose
properties strongly depend on the total mass of the BNS, on the mass ratio and on the 
EOS~\citep[e.g.][]{Hotokezaka:2012ze,Rosswog:2012wb,Bauswein:2013yna,Radice:2016dwd,Bovard:2017mvn,Radice:2018ghv,Radice:2018pdn}.
This ejection is due to tidal torques and shocks
developing at the contact interface between the merging stars, 
when matter is squeezed out by hydrodynamical processes~\cite{Oechslin:2006uk,Hotokezaka:2012ze}.
The expansion of this ejecta component has a velocity of roughly $\vrms \sim0.2\, c$. Moreover, 
this phenomenon generates a distribution of ejected mass denser in the regions across the orbital plane with respect to the region along its orthogonal axis, 
characterized by larger opacities at lower latitudes. In particular, neutrino irradiation (if significant), increases the ejecta $Y_e$ and prevents the formation of lanthanides.
For the anisotropic analyses, the mass profile is taken to be 
$\varrho(\theta)\propto \sin\theta$,
and the opacity profile is take as a step function in the polar angle
characterized by the parameters $(\kappa_{\rm low},\kappa_{\rm high})$,
respectively for low- and high-latitudes,
with a step angle $\theta_{\rm step}=\pi/4$ (see Sec.~\ref{subsec:priors}).
In terms of emitted LC, the described ejecta is characterized 
by a red equatorial component and a blue contribution at higher latitudes.

\paragraph*{Neutrino-driven wind  $({\rm N})$.} 
Simulations of the remnant evolution in the aftermath of a BNS merger reveal the presence of other ejection mechanisms happening over the thermal and viscous evolution timescales~\citep[e.g.][]{Metzger:2008av,Fernandez:2013tya,Perego:2014fma,Perego:2017fho,Decoene:2019eux}. 
If the ejection happens while the remnant is still a relevant source of neutrinos, 
neutrino irradiation has enough time 
to increase $Y_e$ above 0.25, preventing
full $r$-process nucleosynthesis, especially
close to the polar axis.
Detailed simulations~\cite{Perego:2014fma,Martin:2015hxa,Fujibayashi:2017puw,Fujibayashi:2020dvr} show that a relatively small fraction of the expelled disk contributes to this component and its velocity is expected to be $\vrms \lesssim 0.1c$.
For anisotropic analyses, the mass profile is taken to be uniform in the range $\theta\in[0,\pi/3]$ and negligible otherwise,
while the opacity profile is takes as as step function in the polar angle, with a step angle $\theta_{\rm step}=\pi/3$.

\paragraph*{Disk's viscous ejecta  $({\rm V})$.} 
In addition to neutrinos, viscous torques of dynamical and magnetic origin can unbind matter from the disk around massive NSs or black holes~\cite{Metzger:2009xk,Metzger:2014ila,Just:2015fda}.
This viscous component is expected to unbound a large fraction of the disk matter on longer timescale, reaching $\Me\lesssim 10^{-1}\Msun$, with a  relatively low velocity, $\vrms \lesssim 0.05c$. The corresponding ejecta are more uniformly distributed over the polar angle than the dynamical ejecta and the $\nu$-driven wind ejecta. The presence or the lack of a massive NS in the center can influence the $Y_e$ of these ejecta. Then, all angular profiles are assumed to be isotropic for this component~\cite{Wu:2016pnw,Siegel:2017jug}.

We conclude this section by recalling that the main motivation behind the usage of
the semi-analytic model presented above is the optimal compromise between its robustness and
adaptability, essential to model the non-trivial structure of the ejecta, 
and the reduced computational costs, necessary to perform parameter estimation studies.
However, it has been showed that simplified models that avoid the solution of the radiation
transport problem can suffer from systematic uncertainties \cite{Wollaeger:2017ahm}. 
In particular, the analytical model presented in \cite{Grossman:2013lqa}, on which ours is based, 
produces significantly lower light curves.
The comparison with observed kN light curves and more detailed kN models showed how larger
nuclear heating rates $\epsilon_0$ systematically reduce this discrepancy.

\section{Method} 
\label{sec:method}

In this section, we recall the basic concepts of model selection as they are 
stated in the Bayesian theory of probability. Then, we describe 
the statistical technique used for the computations of the Bayes' factors.
As convention, the symbol `$\log$' denotes the natural logarithm 
while a logarithm to a different base is explicitly written when it is used.

\subsection{Model Selection}
\label{subsec:modelselection}

Given some data $d$ and a model $H$ (hypothesis) described by a set of parameters $\bdphi$, the posterior probability is given by the Bayes' theorem:
\be
\label{eq:bayes}
p(\bdphi|d,H)=\frac{p(d|\bdphi,H)\,p(\bdphi|H)}{p(d|H)}\,,
\ee
where $p(d|\bdphi,H)$ is the likelihood function, $p(\bdphi|H)$ is the prior probability
assigned to the parameters and $p(d|H)$ is the evidence. The latter value plays the role of 
normalization constant and it can be computed by marginalizing the likelihood function,
\be
\label{eq:evidence}
p(d|H) = \int_\Theta  p(d|\bdphi,H)\,p(\bdphi|H) {\rm d} \bdphi\,,
\ee
where the integral is computed over the entire parameters' space $\Theta$.

In the framework of Bayesian theory of probability, we can compare two models,
say $A$ and $B$, by computing the ratios of the respective posterior probabilities, also known as {\it Bayes' factor},
\be
\label{eq:bayesfact0}
\B_B^A =\frac{p(A|d,H_A)}{p(B|d,H_B)}\,.
\ee
Using Eq.~\eqref{eq:bayes} we get:
\be
\label{eq:bayesfact1}
\B_B^A =\frac{p(d|A,H_A)}{p(d|B,H_B)}\frac{p(A|H_A)}{p(B|H_B)}=\frac{p(d|A,H_A)}{p(d|B,H_B)}\,,
\ee
where we assumed that the data do not depend on the different hypothesis and that different models are equally likely a priori, i.e. $p(A|H_A)=p(B|H_B)$.
Now suppose that the two models $A,B$ are respectively described by two sets of parameters $\bdphi_A,\bdphi_B$.
Using the marginalization rule we can write:
\be
\label{eq:intZ}
p(d|I,H_I) = \int_{\Theta_I} p(d|\bdphi_I,I,H_I)\,p(\bdphi_I|I,H_I)\,{\rm d}\bdphi_I \,,
\ee
for $I=A,B$. The integral in Eq.~\eqref{eq:intZ} represents the evidence computed for the  
hypotheses $H'_I = \{ H_I , I\}$, for $I=A,B$ (i.e. the involved model becomes 
part of the background hypothesis). Then, we obtain that the Bayes' factor 
$\B_B^A$ can be computed as 
\be
\label{eq:bayesfact2}
\B_B^A =\frac{p(d|H'_A)}{p(d|H'_B)}\,. 
\ee

From the previous results, we understand that if  $\B_B^A>1$ then the model $A$ will be favored
by the data, viceversa if $\B_B^A<1$. It is important to observe that the Bayes' factor implicitly takes 
into account the so called Occam's razor, 
i.e. if two models are both able 
to capture the features of the data,
then the one with lower number of parameters will be favored~\cite{Sivia2006}. 
In our analysis, this is a crucial point since different models have different numbers of parameters.

\subsection{Nested Sampling}
\label{subsec:sampling}

In a realistic scenario, the form of the likelihood function is 
analytically indeterminable and the parameter space has a 
non-trivial number of dimensions.
For these reasons, the estimation of Eq.~\eqref{eq:intZ}
is performed resorting to statistical computational techniques:
we employ the nested sampling Bayesian technique introduced in Ref.~\cite{Skilling:2006} and 
designed to compute the evidence and explore the 
full parameter space. 
The uncertainties associated with the evidence estimations are 
computed according to Ref.~\cite{Skilling:2006}
and increasing the result by one order of magnitude, in order to conservatively take into account 
systematics. 
The latter are expected since the model considered for our analyses (as many others) cannot capture all the physics processes involved in kNe, 
and it suffers of large uncertainties in the atomic physics and radiative processes implementation.

We perform inference with {\tt cpnest}~\cite{cpnest}, a parallelized nested sampling implementation.
We use 1024 live points and, for every step, 
we set a maximum number of 2048 Markov-chain Monte Carlo (MCMC) iterations for the exploration
of the parameter space. The proposal step method used in the MCMC 
is the same as the one implemented as default in {\tt cpnest} software. It corresponds  
to a cycle over four different proposal methods: a random-walk step~\cite{Goodman:2010}, 
a stretch move~\cite{Goodman:2010}, a differential evolution method~\cite{Nelson:2013} 
and a proposal based on the eigenvectors of the covariance matrix of the ensemble samples, 
as implemented in Ref.~\cite{Veitch:2014wba}.

\subsection{Choice of Priors}
\label{subsec:priors}

\begin{table}
	\centering
	\caption{List of intrinsic and extrinsic parameters involved in the analysis
		and the respective prior bounds for the cases of anisotropic geometry. 
		For isotropic geometry cases,
		the bounds are identical except for the opacity $\kappa$ of dynamical
		component (D), where the low-latitude and high-latitude 
		bounds are joined together.}
	\label{tab:params}
	\begin{tabular}{cccccc}
		\hline  \hline
		\multicolumn{6}{c}{Intrinsic Ejecta Parameters $\bdphi_{\rm ej}^{\rm (D,V,N)}$}     \\
		\hline 
		Comp.      & $\Me$                          & $\vrms$                       & $\kappa_{\rm high}$           & $\kappa_{\rm low}$            & $\theta_{\rm step}$      \\
		                 & $[10^{-2}{\rm M}_\odot]$       & $[c]$                             & \multicolumn{2}{c}{$[\igscm]$}                                & $[{\rm rad}]$            \\ \hline
		{D} &{[}0.1, 10{]} & {[}0.15,0.333{]} & {[}0.1,5{]} & {[}5,30{]}   & {$\pi/4$} \\
		{N} & {{[}0.01,0.75{]}}  & {{[}0.05,0.15{]}}  & 
		\multicolumn{2}{c}{{{[}0.01,5{]}} }& {$\pi/3$} \\
		 {V} & {{[}1,20{]}}    & {{[}0.001,0.1{]}}  &\multicolumn{2}{c}{{[}0.01,30{]}}       & {--}      \\\hline  \hline
	\end{tabular} \\
	\vspace{0.5cm}
	\begin{tabular}{ccc}
		\hline\hline
		\multicolumn{3}{c}{Intrinsic Global Parameter $\bdphi_{\rm glob}$}                                                     \\ \hline
		$\Tfni$ & {[}K{]}                          & {[}500, 8000{]}                      \\
		$\Tfla$ & {[}K{]}                          & {[}500, 8000{]}                      \\
		$\eps_0$                 & [$\ergsg$] & $[2\times 10^{17}, 5\times 10^{19}]$ \\\hline\hline
		  \end{tabular}\\
	\vspace{0.5cm}
		\begin{tabular}{ccc}
			\hline\hline                                    
		\multicolumn{3}{c}{Extrinsic Parameters $\bdphi_{\rm ext}$}                                                           \\ \hline
		$D_L$                    & {[}Mpc{]}                        & {[}15,50{]}                          \\
		$\iota$                 & {[}deg{]}                       & {[}0,70{]}             \\\hline\hline             
        \end{tabular}

\end{table}

In our analysis we assume the sky position of the source 
to be known
and the time of coalesce to be the same of the trigger time of GW170817~\cite{TheLIGOScientific:2017qsa}.
Furthermore, we do not take into account the redshift contribution,
given the larger systematic uncertainties in the model.
We employ the parameters shown in Tab.~\ref{tab:params},
that can be divided in three subsets:
the intrinsic ejecta parameters $\bdphi_{\rm ej}^{({\rm D}, {\rm V}, {\rm N})}$,
the intrinsic global parameters $\bdphi_{\rm glob}$,
and the extrinsic parameters $\bdphi_{\rm ext}$ .

The intrinsic ejecta parameters, $\bdphi_{\rm ej}^{(k)}$ for $k={\rm D}, {\rm V}, {\rm N}$, characterize the properties of each ejecta component and they are the amount of ejected mass, $\Me$,
the rms velocity of the fluid, $\vrms$, and their grey opacity, $\kappa$. 
Under the assumption of isotropic geometry, 
the intrinsic ejecta parameters $\bdphi_{\rm ej}^{(k)}$ are defined by a single value 
for every shell, i.e. a single number characterizes the entire profile of the 
parameter of interest, since it is spherically symmetric.
However, for anisotropic cases, we have to introduce more than one independent parameters
to describe an angular profile for a specific variable: this is the case of the opacity parameter of the dynamical component, where the profile is chosen as step functions characterized by two different parameters, $\kappa_{\rm low}$ and  $\kappa_{\rm high}$, respectively at low 
and high latitudes. In such a cases, the angle $\theta_{\rm step}$ is introduced to denote
the angle at which the profile changes value, as mentioned in Sec.~\ref{subsec:multicomponents}.

The intrinsic global parameters,
$\bdphi_{\rm glob}$, represent the properties of the source common to every component, such as the floor temperatures, $\Tfni$ and $\Tfla$, 
and the heating rate constant $\eps_0$. In principle, the latter is a universal property
which defines the nuclear heating rate as expressed in Eq.~\ref{eq:epsnuc}.
The whole set of intrinsic parameters, $\bdphi_{\rm glob}$ and $\bdphi_{\rm ej}^{(k)}$,
determines the physical dynamics of the system and, therefore, they determine the properties of the kN emission, irrespectively of the observer location.
 
The extrinsic parameters, $\bdphi_{\rm ext}$, are the luminosity distance of the source, $D_L$
and the viewing angle $\iota$.
These parameters do not depend on the
physical properties of the source and they are related with the 
observed signal through geometrical argumentation. 

The prior distributions for all the parameters are taken uniform in their bounds, 
except for the followings.
For the extrinsic parameters $\bdphi_{\rm ext}=\{D_L,\iota\}$,
we set the priors equal to the marginalized posterior distributions
coming from the low-spin-prior measurement of GW170817~\cite{LIGOScientific:2018mvr};
For the heating rate factor $\eps_0$, we use a uniform prior distribution in $\log\eps_0$, i.e. $p(\eps_0|H)\propto {\eps_0}^{-1}$, 
since this parameter strongly affects the LC and it is free to vary in a wide range. Moreover, we adopt a prior range
according with the estimation given in Ref.~\cite{Korobkin:2012uy}.
Tab.~\ref{tab:params} shows the prior bounds used for the analysis of the anisotropic cases.
For the isotropic studies, 
the bounds are identical except for the opacity $\kappa$ of dynamical
component, where the low-latitude and high-latitude 
bounds are joined together. 

\subsection{Likelihood Function}
\label{subsec:likelihood}

The data $\left\{ d_{b,i} \pm \sigma_{b,i} \right\}$ are the apparent magnitudes observed from AT2017gfo, with their standard deviations. They have been collected from~\cite{Villar:2017wcc}, where all the precise reference to the original works and to the data reduction techniques can be found. The index $b$ runs over all considered photometric bands, covering a wide photometric range from the UV to the NIR, while for each band $b$ the index $i$ runs over the corresponding sequence of $N_b$ temporal observations.
Additionally, the magnitudes have been corrected for Galactic extinction~\cite{Cardelli.etal:1989}.
We introduce a Gaussian likelihood function in the apparent magnitudes
with mean and variance, $d_{b,i}$, $\sigma^2_{b,i}$, from 
the observations of AT2017gfo, 
\be
\label{eq:likelihood}
{\log p(\bdphi|d, H)} \propto -\frac{1}{2} \sum_{b} \sum_{i=1}^{N_b}\frac{\left| d_{b,i} - \mg_{b,i}(\bdphi)\right|^2}{\sigma^2_{b,i}} \,,
\ee
where $\mg_{b,i}(\bdphi)$ are the magnitudes generated by
the LC model,
of Sec.~\ref{sec:knmodel},
which encodes the dependency on the parameters $\bdphi$,
for every band $b$ at different times $i$.
The likelihood definition Eq.~\eqref{eq:likelihood} 
is in accordance with the residuals introduced in Ref.~\cite{Perego:2017wtu}
and it takes into account the uncertainties due to possible technical issues 
of the instruments and generic non-stationary contributions, 
providing a good characterization of the noise
\footnote{Also the work presented in Ref.~\cite{Villar:2017wcc}
employs a Gaussian likelihood, with the inclusion of an additional uncertainty parameter;
while, in Ref.~\cite{Coughlin:2017ydf}, the authors proposed
a likelihood distributed as a $\chi^2$.}.
For both geometric configurations, isotropic (ISO) and anisotropic (ANI), 
we perform Bayesian analyses using 
different combinations of components,
testing the capability to fit the data.


\section{Results} 
\label{sec:results}

In this section we present the results gathered from the Bayesian analysis.
In Sec.~\ref{sec:lcfit} we describe the capability of the synthetic LCs to fit the observed data.
After that, in Sec.~\ref{sec:evidence}, we discuss the estimated evidence inferring the preferred model.
Finally, in Sec.~\ref{sec:posterior}, we discuss the interpretation of the recovered posterior distributions.

\subsection{Light Curves} 
\label{sec:lcfit}

\begin{figure*}
	\centering 
	\includegraphics[width=\textwidth]{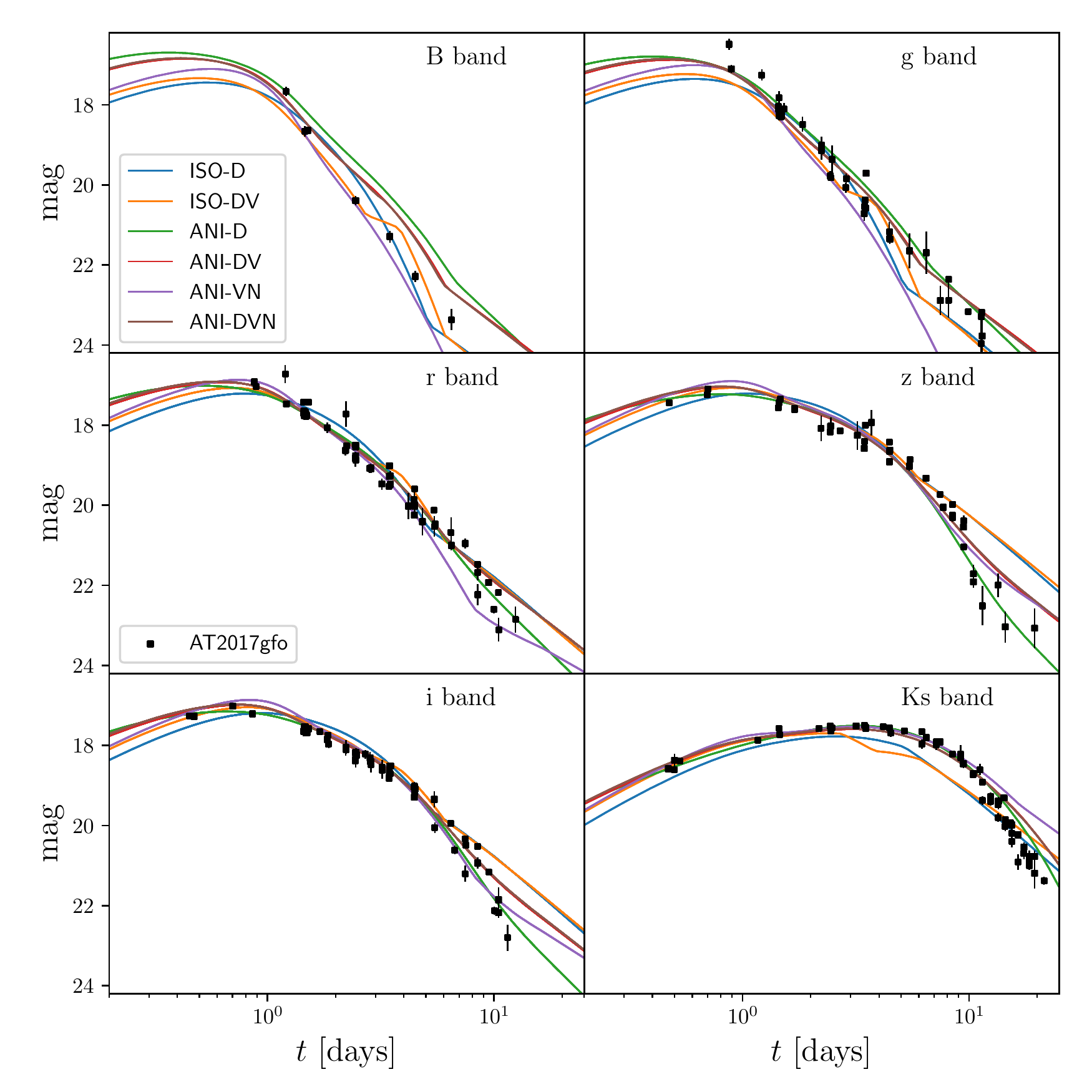}
	\caption{Apparent magnitudes 
					computed using the maximum-likelihood parameter
					for each considered model;
					ISO-D in blue,
					ISO-DV in yellow,
					ANI-D in green,
					ANI-DV in red,
					ANI-VN in purple and 
					ANI-DVN in brown.
					The different panels refer to different photometric bands,
					respectively $B$, $g$, $r$, $z$, $i$ and $Ks$.
					The black squares are the observed data of AT2017gfo for the corresponding 
					photometric band with the respective standard deviations.}
	\label{fig:lcs}
\end{figure*}

\begin{figure*}
	\centering 
	\includegraphics[width=\textwidth]{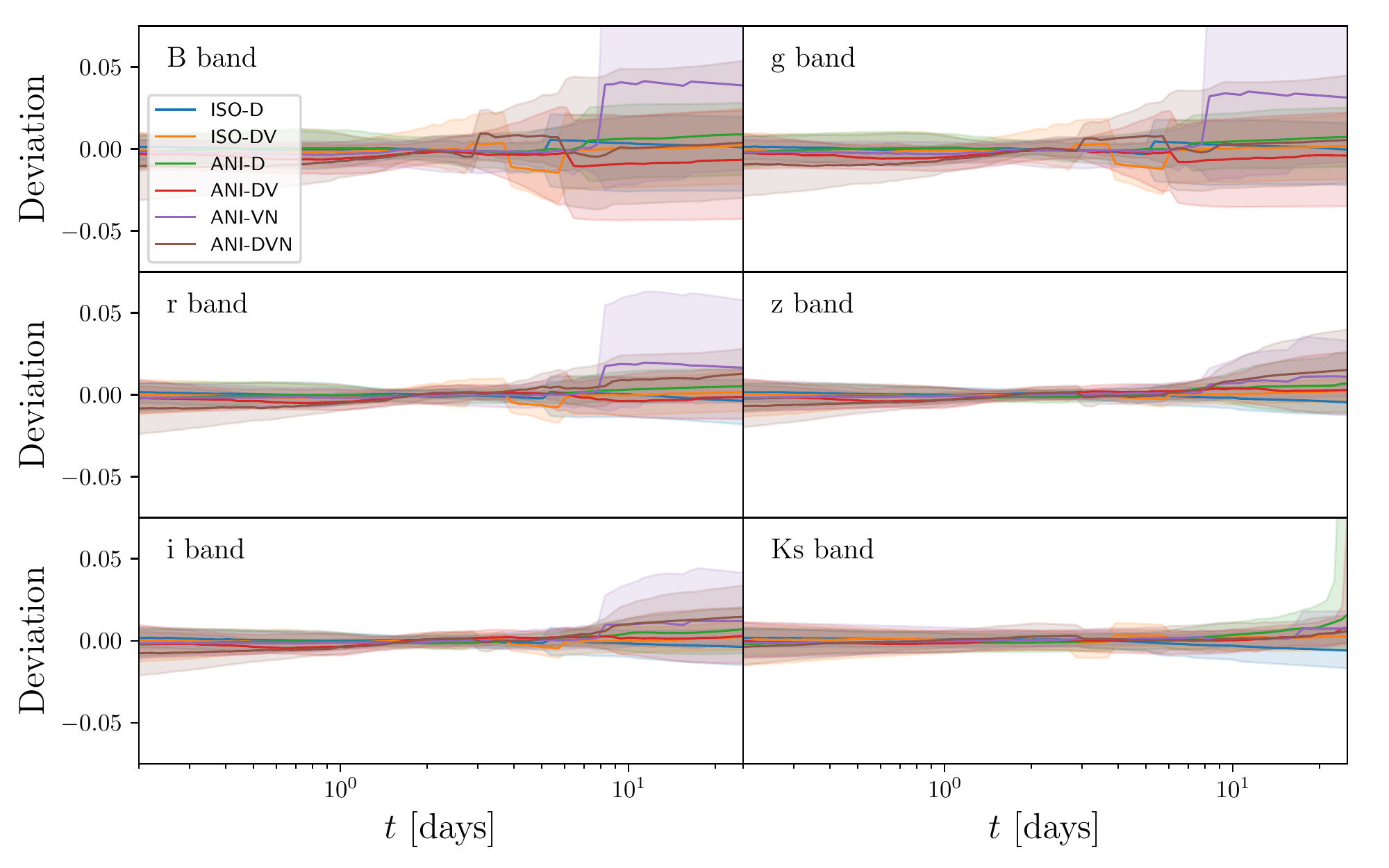}
	\caption{Deviations from the maximum-likelihood template
					of the LCs computed from the whole set of posterior samples.
					The solid lines represent the median values and the shadowed areas are 
					the 90\% credible regions.
					Different color refers to a different model; respectively, 
					ISO-D in blue,
					ISO-DV in yellow,
					ANI-D in green,
					ANI-DV in red,
					ANI-VN in purple and 
					ANI-DVN in brown.
					The different panels show different photometric bands,
					respectively $B$, $g$, $r$, $z$, $i$ and $Ks$.}
	\label{fig:lcerr}
\end{figure*}

Figure~\ref{fig:lcs} shows the LCs computed from the recovered maximum-likelihood 
parameters for each discussed model. The estimated LCs are compared with AT2017gfo 
data for six representative photometric bands. 
Moreover, Fig.~\ref{fig:lcerr} shows the uncertainties associated
with the estimated LCs, computed over the recovered posterior samples, for each considered model.
Generally, the errors associated with the near UV (NUV) magnitudes
are larger compared with the other bands, reflecting the 
lower number of data points in this photometric region.
Furthermore, none of the considered model is able to fully capture the trend 
described by the observed data in the Ks band for time larger then 10~days,
within the provided prior bounds.
This is expected from the simplified treatment 
of the radiation transport and the approximated
heating rate in our models.

The isotropic models (ISO-D and ISO-DV) give a good fitting 
to the data for early times
and their LCs capture the general trends of the data.
However, for times larger than ${\sim}8$~days, these models do not capture all the features of the data within the provided prior bounds.
This inaccuracy is particularly evident in the NIR, 
where the LCs predicted by the ISO-D and the ISO-DV
models do not recover the correct slopes of the data.

The anisotropic single-component case, ANI-D,
is apt at adapting the model to the different features present in the data,
even for large time-scales. 
However, it overestimates the kN emission in the blue band. 
This inconsistency could be reduced allowing the high latitude opacity
parameter $\khigh$ to lower values.
Regarding the anisotropic two-components models, the ANI-VN 
gives a good fitting for early times, but the model largely underestimates the data 
at times ${\gtrsim}5$~days. This is due to the absence of a fast blue component. 
The anisotropic ANI-DV model gives LCs similar to ANI-D except for 
a slight excess of power for time ${\gtrsim}10$~days, especially in the NIR region, 
i.e. $z$, $K$ and $K_s$ bands.
This behavior could be mitigated by reducing the lower bound on the $\Tfla$ parameter.
However, it could also indicate a significant deviation from the black-body emission
adopted in our model at late times.
Furthermore, the ANI-DV model overshoots the data in the NUV,
as it is for the respective single-component case ANI-D.
This can be explained looking at the recovered value of 
dynamical ejected mass, which exceeds theoretical expectations 
estimated from NR simulations~\cite{Perego:2019adq,Endrizzi:2019trv,Nedora:2019jhl,Bernuzzi:2020txg,Nedora:2020pak}(see Sec.~\ref{sec:ani-dv}).
Similar considerations hold for the anisotropic three-component case ANI-DVN.
However, the uncertainties on the estimated LCs for this model are narrower with respect to
the ones obtained from the ANI-DV, corresponding to an improvement in the 
capability of constraining the measurement.
The main improvement of the three-component ANI-DVN model over the two-component ANI-DV
model lies in its ability to better fit early-times data
due to the inclusion of a third component.

\subsection{Evidences} 
\label{sec:evidence}

\begin{table}
	\centering    
	\caption{Estimated log-evidences for the analyzed kNe models.
		The reported uncertainties refers to the standard deviations
		estimated according to Ref.~\cite{Skilling:2006}.}
	\begin{tabular}{ccc}        
		\hline\hline
		Profile & Components  &$\log p(d|{\rm Model})$\\
		\hline
		\hline
		ISO & D &$-23510\pm 1$ \\
		\hline
		ISO & D+V &$ -19719 \pm 1$ \\
		\hline
		\hline
		ANI & D &$-9920\pm 1 $\\ 
		\hline   
		ANI & N+V &$-11103 \pm 1$ \\
		\hline
		ANI & D+V& $-9556 \pm 1$ \\ 
		\hline
		ANI & D+N+V & $-9439 \pm 1$ \\ 
		\hline\hline
	\end{tabular}
	\label{tab:evidence}
\end{table}

The logarithmic evidences estimated for the considered models 
are shown in Tab.~\ref{tab:evidence}.
The evidence increases with the number of models' components.
This is consistent with the hierarchy observed in the LC residuals, and the better match to the data for multi-component models.
The only exception is the ANI-NV case, 
for which
the features of the data at late times are not well captured due to the absence of a fast equatorial component.
Furthermore, for a fixed number of components, the anisotropic geometries are always favored with respect to isotropic geometries,
with a $\log \B_{\rm ISO}^{\rm ANI}$ of the order of $10^{4}$. 
The preferred model among the considered cases is the anisotropic three-component,
in agreement with previous findings, e.g.~\cite{Cowperthwaite:2017dyu, Perego:2017wtu, Villar:2017wcc}.

\subsection{Posterior Distributions} 
\label{sec:posterior}

\begin{table*}
	\centering    
	\caption{Recovered values from the posterior distributions of the 
		of the intrinsic ejecta parameters. The reported quantities are the means 
		with the 90\% credible regions.
		The conventions $\gtrsim$, $\lesssim$ denote 
		marginalized posterior distributions constrained respectively 
		around the upper and the lower prior bounds.
		We remark that $\klow$ and $\khigh$ refer respectively to the gray opacity parameters for low and high latitudes.}
	\resizebox{0.99\textwidth}{!}{
		\begin{tabular}{c|cccc|cccc|ccc}        
			\hline\hline
			Model &\multicolumn{4}{c|}{ Dynamical ejecta}  & \multicolumn{4}{c|}{ Viscous ejecta} & \multicolumn{3}{c}{$\nu$-driven wind}\\
			& $\Me$ & $\vrms$ & $\khigh$ & $\klow$& $\Me$ & $\vrms$ & \multicolumn{2}{c|}{ $\kappa$}& $\Me$ & $\vrms$ & $\kappa$\\
			& $\left[10^{-2} \Mo\right]$ & $[c]$ & \multicolumn{2}{c|}{ $\left[{\rm cm}^2\,{\rm g}^{-1}\right]$}& $\left[10^{-2} \Mo\right]$ & $[c]$ & \multicolumn{2}{c|}{ $\left[{\rm cm}^2\,{\rm g}^{-1}\right]$}& $\left[10^{-2} \Mo\right]$ & $[c]$ &  $\left[{\rm cm}^2\,{\rm g}^{-1}\right]$\\
			\hline
			\hline
			ISO-D 
			&$0.787^{+0.016}_{-0.017}$
			& $0.1758^{+ 0.0007}_{-0.0008}$
			&\multicolumn{2}{c|}{$6.14^{+ 0.11}_{-0.10}$}
			& -- & -- & -- & --
			& -- & -- & -- 
			\\ 
			ISO-DV 
			&$1.139^{+0.048}_{-0.044}$
			& $0.213^{+ 0.003}_{-0.003}$
			&\multicolumn{2}{c|}{$4.13^{+ 0.08}_{-0.09}$}
			&${\lesssim}1$
			& ${\gtrsim}0.1$
			&\multicolumn{2}{c|}{$4.99^{+ 0.12}_{-0.11}$}
			& -- & -- & -- 
			\\
			\hline
			ANI-D 
			&$0.807^{+0.022}_{-0.018}$
			& $0.236^{+ 0.001}_{-0.002}$
			&${\lesssim}0.1$
			& ${\gtrsim}30$
			& -- & -- & -- & --
			& -- & -- & -- 
			\\ 
			ANI-DV
			&$1.231^{+0.041}_{-0.048}$
			& $0.233^{+ 0.002}_{-0.002}$
			&${\lesssim}0.1$
			& $12.3^{+ 0.6}_{-0.5}$
			&${\lesssim}1$
			& $ 0.0276^{+ 0.0007}_{-0.0006}$
			&\multicolumn{2}{c|}{$2.23^{+ 0.05}_{-0.05}$}
			& -- & -- & -- 
			\\ 
			ANI-VN 
			& -- & -- & -- & --
			&${\lesssim}1$
			& $ 0.0064^{+ 0.0001}_{-0.0001}$
			&\multicolumn{2}{c|}{$0.45^{+ 0.01}_{-0.01}$}
			& $\gtrsim 0.75$
			& $  0.0998^{+ 0.0003}_{-0.0008}$
			& $  1.002^{+ 0.006}_{-0.002}$
			\\
			ANI-DVN
			&$1.378^{+0.063}_{-0.071}$
			& $0.233^{+ 0.002}_{-0.002}$
			&${\lesssim}0.1$
			& $11.1^{+ 0.7}_{-0.6}$
			&${\lesssim}1$
			& $ 0.0318^{+ 0.0008}_{-0.0008}$
			&\multicolumn{2}{c|}{$2.96^{+ 0.07}_{-0.09}$}
			& $0.247^{+0.025}_{-0.061}$ 
			&  $0.0502^{+0.0006}_{- 0.0002}$
			&  $2.29^{+0.14}_{- 0.09}$
			\\ 
			\hline\hline
		\end{tabular}
	}
	\label{tab:intr_posteriors}
\end{table*}

\begin{table}
	\centering    
	\caption{Recovered values from the posterior distributions of the 
		of the global intrinsic parameters and of the extrinsic parameters. 
		The reported quantities are the means 
		with the 90\% credible regions.
		The conventions $\gtrsim$, $\lesssim$ denote 
		marginalized posterior distributions constrained respectively 
		around the upper and the lower prior bounds.}
	\resizebox{0.49\textwidth}{!}{
		\begin{tabular}{c|ccccc}        
			\hline\hline
			Model & $\Tfni$ & $\Tfla$ & $\eps_0$ & $\iota$ & $D_L$\\
			& $\left[{\rm K}\right]$ 
			& $\left[{\rm K}\right]$  
			& $\left[10^{18}{\rm erg} \, {\rm g}^{-1}\, {\rm s}^{-1}\right]$ 
			&  $\left[{\rm deg}\right]$ 
			& $\left[{\rm Mpc}\right]$ \\
			\hline\hline
			ISO-D 
			& ${4335}^{+3157}_{-3427}$ 
			& ${2484}^{+450}_{-410}$ 
			& ${66.5}^{+1.5}_{-1.4}$ 
			& ${33}^{+27}_{-25}$ 
			& ${\gtrsim}50$ 
			\\
			ISO-DV & ${6740}^{+778}_{-612}$ & ${1126}^{+243}_{-311}$ & ${21.21}^{+0.05}_{-0.05}$ & ${34}^{+24}_{-26}$ & ${48.5}^{+0.3}_{-0.4}$ \\
			\hline
			ANI-D 
			& ${5064}^{+47}_{-50}$ 
			& ${746}^{+219}_{-223}$ 
			& ${161}^{+3}_{-5}$ 
			& ${43.9}^{+0.5}_{-0.5}$ 
			& ${\gtrsim}50$  
			\\
			ANI-DV 
			& ${5031}^{+105}_{-99}$ 
			& ${704}^{+175}_{-180}$ 
			& ${38.7}^{+0.9}_{-0.9}$ 
			& ${43.9}^{+0.5}_{-0.5}$ 
			& ${\gtrsim}50$  
			\\
			ANI-VN 
			& ${3356}^{+56}_{-35}$ 
			& ${\lesssim}{500}$ 
			& ${8.5}^{+0.1}_{-0.1}$ 
			& ${52}^{+1}_{-1}$ 
			& ${22.6}^{+0.2}_{-0.2}$ \\
			ANI-DVN 
			& ${5995}^{+105}_{-118}$ 
			& ${\lesssim}{500}$ 
			& ${30.4}^{+0.2}_{-0.1}$ 
			& ${57}^{+1}_{-1}$ 
			& ${\gtrsim}50$  
			\\
			\hline\hline
		\end{tabular}
	}
	\label{tab:extr_posteriors}
\end{table}

In the following paragraphs, we discuss the properties of the posterior distributions for each model and their physical interpretation.
Table~\ref{tab:intr_posteriors} and Tab.~\ref{tab:extr_posteriors} show
the mean values of the parameters, and their 90\% credible regions, extracted from the 
recovered posterior distributions.
A general fact is that the marginalized posterior for the ejected mass of the viscous component
is always constrained against the lower bound $10^{-2}~\Mo$, when this component is involved.
Moreover, for the majority of the analyses, the distance parameter is biased towards larger values,
inconsistently with the estimates from Ref.~\cite{TheLIGOScientific:2017qsa,GBM:2017lvd},
and the heating rate parameter $\eps_0$ is generally overestimated 
comparing with the estimates from nuclear calculations~\cite{Korobkin:2012uy,Barnes:2016umi, Kasen:2018drm,
  Barnes:2020nfi, Zhu:2020eyk}.
 This behavior can be explained from Eqs.~\eqref{eq:epsnuc}, \eqref{eq:spectral_flux} and~\eqref{eq:mag}:
$D_L$ and $\eps_0$ are largely degenerate
and both concur to determine the brightness of the 
observed LCs. Thus, the correlations between these parameters
induce biases in the recovered values.
The physical explanation of this effect can be motivated 
with the poor characterization of the model in the NIR bands:
this lack of knowledge generates a fainter kN in this
photometric region
and, in order to match the observed data, the 
recovered heating rate are larger.
Note that this bias concurs in 
the overestimation of the LC in the high-frequency bands (i.e. $U$, $B$ and $V$),
where the number of measurements is lower with respect to the other 
employed bands.

\subsubsection{ISO-D}
\label{sec:iso-d}

We start considering the simplest employed model,
the isotropic one-component model labelled as ISO-D.
Fig.~\ref{fig:m_v_dyn} shows the marginalized posterior distribution
in the $(\Me,\vrms)$ plane.
The velocity is constrained around ${\sim}0.18\,c$
while the ejected mass lies around $8 {\times} 10^{-3}~\Mo$,
both in agreement with the observational
results recovered in Ref.~\cite{Villar:2017wcc,  Cowperthwaite:2017dyu,Abbott:2017wuw,Coughlin:2018miv}.
Moreover, the opacity posterior peaks in proximity of $\kappa\sim6~\igscm$,
consistently with Ref.~\cite{Cowperthwaite:2017dyu}.

Regarding the extrinsic parameters, the posterior for the 
inclination angle $\iota$ is coincident with the imposed prior,
since the employed profiles do not depend on this coordinate.
The model is not able to constrain the value of $\Tfni$, which returns 
a posterior identical to the prior, while $\Tfla$ is recovered around $2500~{\rm K}$.
The obtained flat posterior distribution for the $\Tfni$ parameter highlights 
the unsuitability of this model in capturing the features of the observed data.

\subsubsection{ANI-D}
\label{sec:ani-d}

\begin{figure}
	\centering 
	\includegraphics[width=0.48\textwidth]{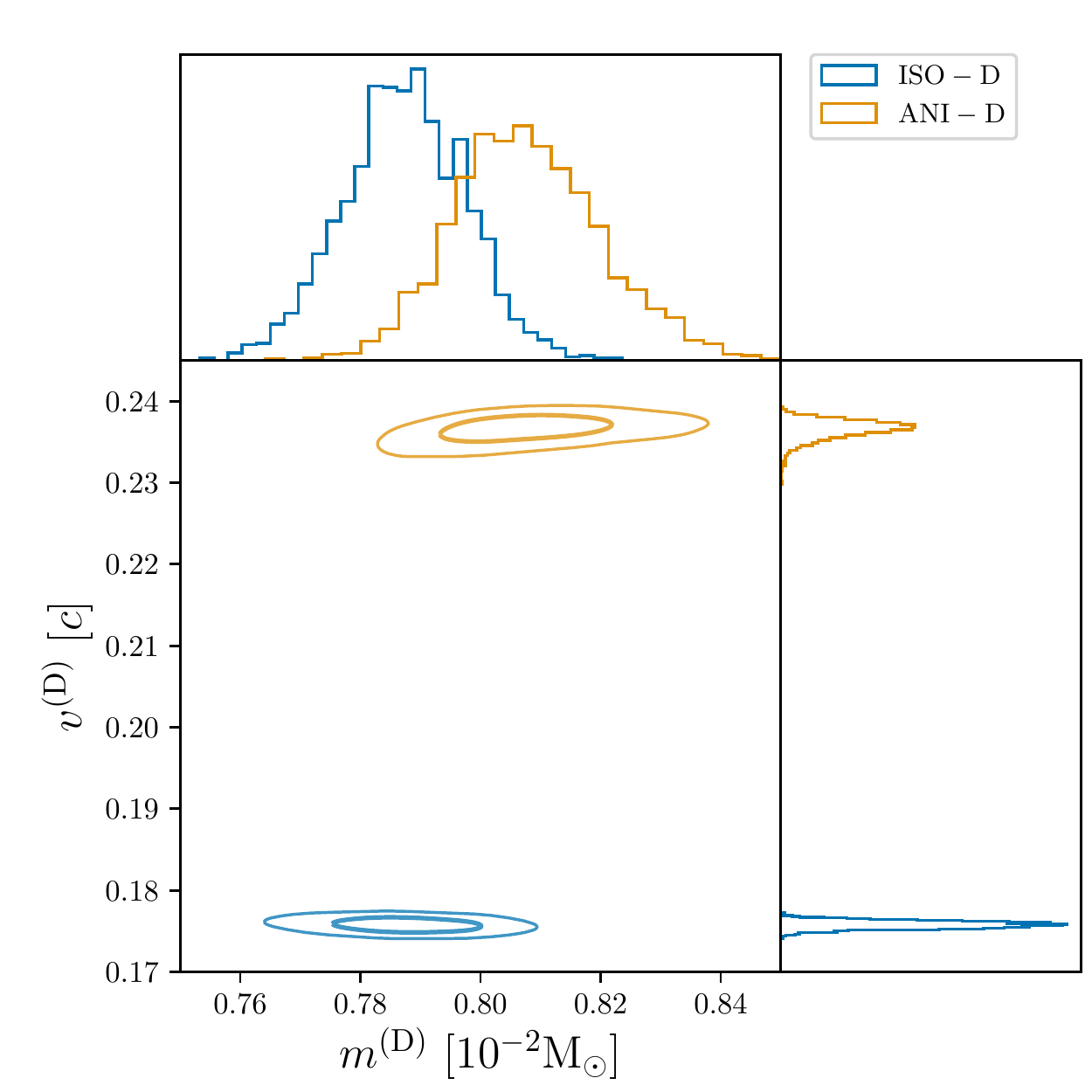}
	\caption{Marginalized posterior distribution 
		of ejected mass $\Me$ and velocity 
		$\vrms$ of dynamical component 
		for the one-component studies, ISO-D and ANI-D.
		The anisotropic case requires larger velocities in order to 
		fit the observed data.}
	\label{fig:m_v_dyn}
\end{figure}

For the anistropic single-component model ANI-D,
the value of the ejected mass agrees with the one coming from the ISO-D case.
However, in order to fit the data, ANI-D requires a larger velocity, ${\sim}0.23\,c$,
as shown in Fig.~\ref{fig:m_v_dyn}.
The high-latitude opacity is constrained around the lower bound $0.1~\igscm$
while the low-latitude contribution exceeds above $30~\igscm$,
that largely differs from the respective isotropic case, ISO-D. 
In practice, that is due to the lack of ejected mass that is 
balanced with a more opaque environment.
Nevertheless, according to the estimated evidences, this model is preferred with
respect to the isotropic case. The reason is clear from Fig.~\ref{fig:lcs}:
the anisotropic model is able to characterize the late-times features of the data.
	The heating rate parameter $\eps_0$ is largely biased towards larger values 
	with respect to the results of Ref.~\cite{Korobkin:2012uy},
	in order to compensate the lack of ejected matter. 
	Indeed, a larger heating factor $\eps_0$ leads to brighter LCs,
	and this effect is capable to mimic an increase in the amount of ejected matter.

The posterior distribution for viewing angle $\iota$ peaks around 44~degrees,
inconsistently with the estimations coming from the GRB analysis~\cite{Monitor:2017mdv,Savchenko:2017ffs,Ghirlanda:2018uyx}.
Moreover, unlike the ISO-D case, both temperature parameters $\Tfni$ and $\Tfla$
are well constrained for the ANI-D analysis: these parameters affect mostly the late-times model, 
modifying the slope of the recovered LCs.
Thus, these terms are
 responsible for the
improvement in the fitted LCs.

\subsubsection{ISO-DV}
\label{sec:iso-dv}

Figure~\ref{fig:23comp_intr} shows the posterior distribution 
for some exemplary intrinsic ejecta parameters.
For both components, the individual most-likely value for ejected mass parameter lies 
around ${\sim10^{-2}~\Mo}$, 
in agreement with the measurement presented in Ref.~\cite{Abbott:2017wuw}.
This range of values is slightly overestimating the expectations coming 
from NR simulations for the dynamical component~\cite{Perego:2019adq,Nedora:2019jhl,Endrizzi:2019trv,Nedora:2020pak,Bernuzzi:2020txg}.
This could be explained by considering the effect of the spiral-wave wind~\cite{Nedora:2019jhl}, that constitute a massive and fast ejecta on
timescales of $10-100$~ms. The spiral-wave wind is not considered as
components in our models because it would be highly degenerate with
the dynamical ejecta.
The recovered opacity parameters are roughly $4{-}5~\igscm$.
The velocity of the dynamical component is greater than secular velocity,
accordingly with the theoretical expectations.
Comparing with other fitting models,
the recovered ejected masses $\Me^{\rm (D)}$
result smaller with respect to the analogous analysis of Ref.~\cite{Villar:2017wcc}, while the results roughly agree with the estimations 
coming from Ref.~\cite{Coughlin:2018miv}.
However, it is not possible to perform an apple-to-apple comparison 
between these results, due to the systematic
differences in modeling between the semi-analytical model (used in this work)
and the radiative-transport methods employed in Ref.~\cite{Villar:2017wcc,Coughlin:2018miv}.
	
The temperature parameters, $\Tfni$ and $\Tfla$, are much more constrained comparing
with the respective isotropic single component case ISO-D,
and this is reflected in the improvement of fitting the different trends of the data in the high-frequency bands.
The marginalized posterior distribution of the inclination angle is coincident with 
the prior, according with the isotropic description. 
Furthermore, 
the biases on \oldtxt{the distance $D_L$ and} the heating parameter $\eps_0$
are reduced with respect to the ISO-D, since
two-component case accounts for a larger amount of total ejected mass.
  Indeed, increasing the number of ejecta components other than the dynamical one, 
  the overall kN becomes brighter since additional terms, becoming transparent at larger times, 
  are included into the computation of the emitted flux. Then, $\eps_0$ tends
  towards lower values in order to compensate this effect and fit the data.
According with the estimated evidences, the isotropic two-components ISO-DV
model is disfavored with respect to the anisotropic single-component ANI-D.
The main difficulty of ISO-DV is, again, to
fit the data at late-times. 

\subsubsection{ANI-DV}
\label{sec:ani-dv}

\begin{figure*}
	\centering 
	\includegraphics[width=0.99\textwidth]{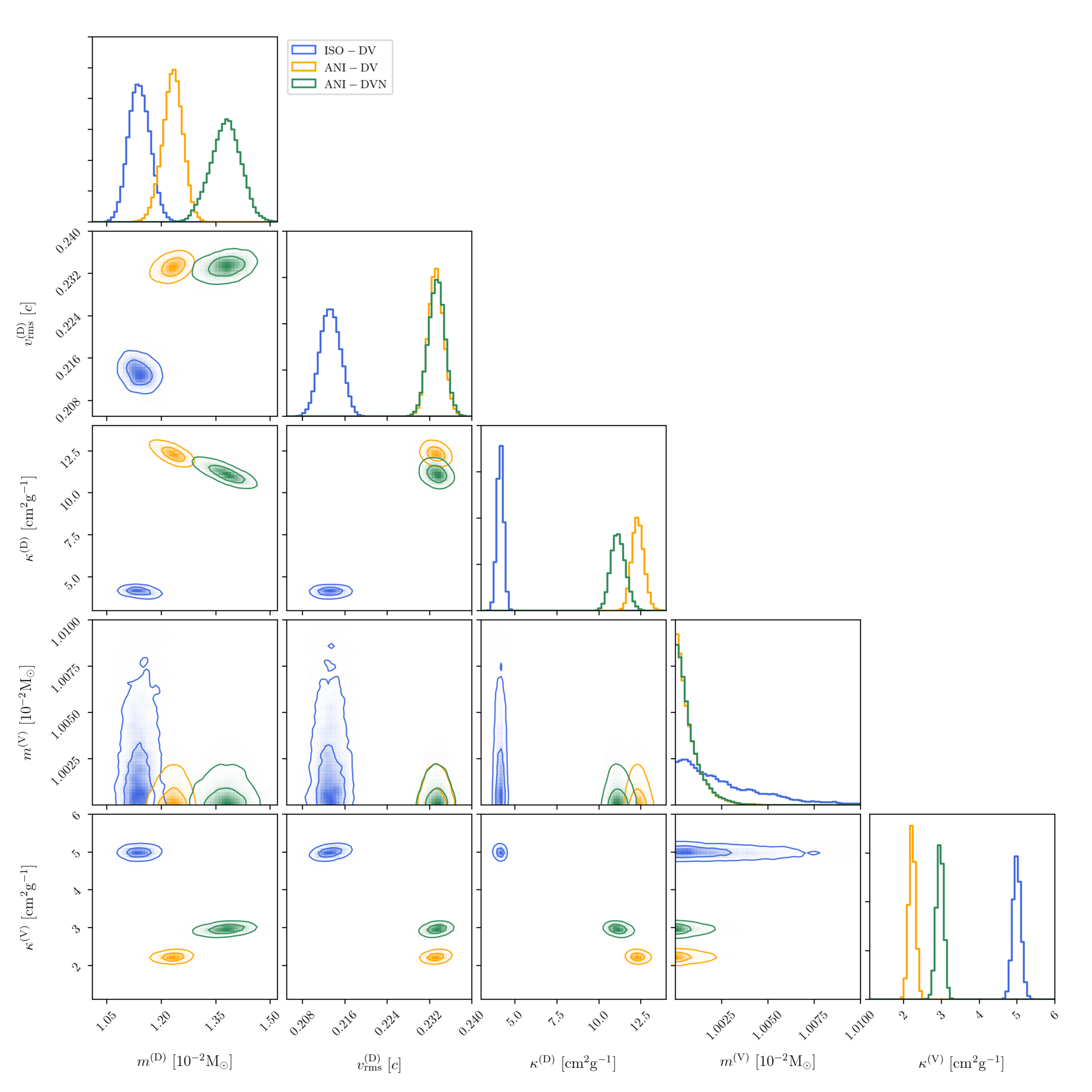}
	\caption{Marginalized posterior distribution 
		for some exemplary ejecta intrinsic parameters 
		extracted from the analysis of ISO-DV, ANI-DV and ANI-DVN.
		The reported parameters are the ejected mass $m^{\rm (D)}$, the velocity $\vrms^{\rm (D)}$ and the low-latitude opacity $\klow^{\rm (D)}$ for the dynamical component, while for the viscous component, we report the ejected mass $m^{\rm (V)}$ and the opacity $\klow^{\rm (V)}$.
		For ISO-DV, the low-latitude opacity of the dynamical component
		is replaced with the overall opacity $\kappa^{\rm (D)}$, due to the 
		different geometry.} 
	\label{fig:23comp_intr}
\end{figure*}

The ANI-DV model is the second best fitting model to AT2017gfo among the 
considered cases.
Fig.~\ref{fig:23comp_intr} shows the posterior distribution 
for some exemplary intrinsic parameters 
of the dynamical and the viscous components.
The ejected mass value lies around ${\sim}10^{-2}~\Mo$,
in agreement with previous estimates~\cite{Abbott:2017wuw}.
On the other hand, the recovered
mass slightly overestimates the 
results coming from targeted NR
simulations~\cite{Perego:2019adq,Nedora:2019jhl,Endrizzi:2019trv,Nedora:2020pak,Bernuzzi:2020txg}, similarly to ISO-DV (see Sec.~\ref{sec:iso-dv}).
The velocity is well constrained around ${\sim}0.23\,c$.
The recovered low-latitude opacity corresponds roughly to $12~\igscm$
and high-latitude opacity is constrained around the lower bound, $0.1~\igscm$.
This result can be explained by considering that the mass of the dynamical component slightly overshoots the 
NR expectations~\cite{Perego:2019adq,Nedora:2019jhl,Endrizzi:2019trv,Nedora:2020pak,Bernuzzi:2020txg}
(of a factor ${\sim}1.25$), and by
noticing that the ejected mass correlates with the luminosity distance 
and the heating factor (that are generally biased).
This combination generates the overestimation of the data in the NUV region. 
In order to improve the fitting to the observed data, the 
model tries to compensate this effect and the
high-latitude opacity tends to move towards lower values.

Concerning the viscous component,
its velocity results an order of magnitude smaller than the one of the
dynamical ejecta, in agreement with the expectations. This enforce the hypothesis 
for which the viscous ejecta contributes mostly to the red kN.
The posterior distribution of opacity parameter peaks around ${\sim}5~\igscm$,
denoting a medium opaque environment.

\begin{figure*}
	\centering 
	\includegraphics[width=0.99\textwidth]{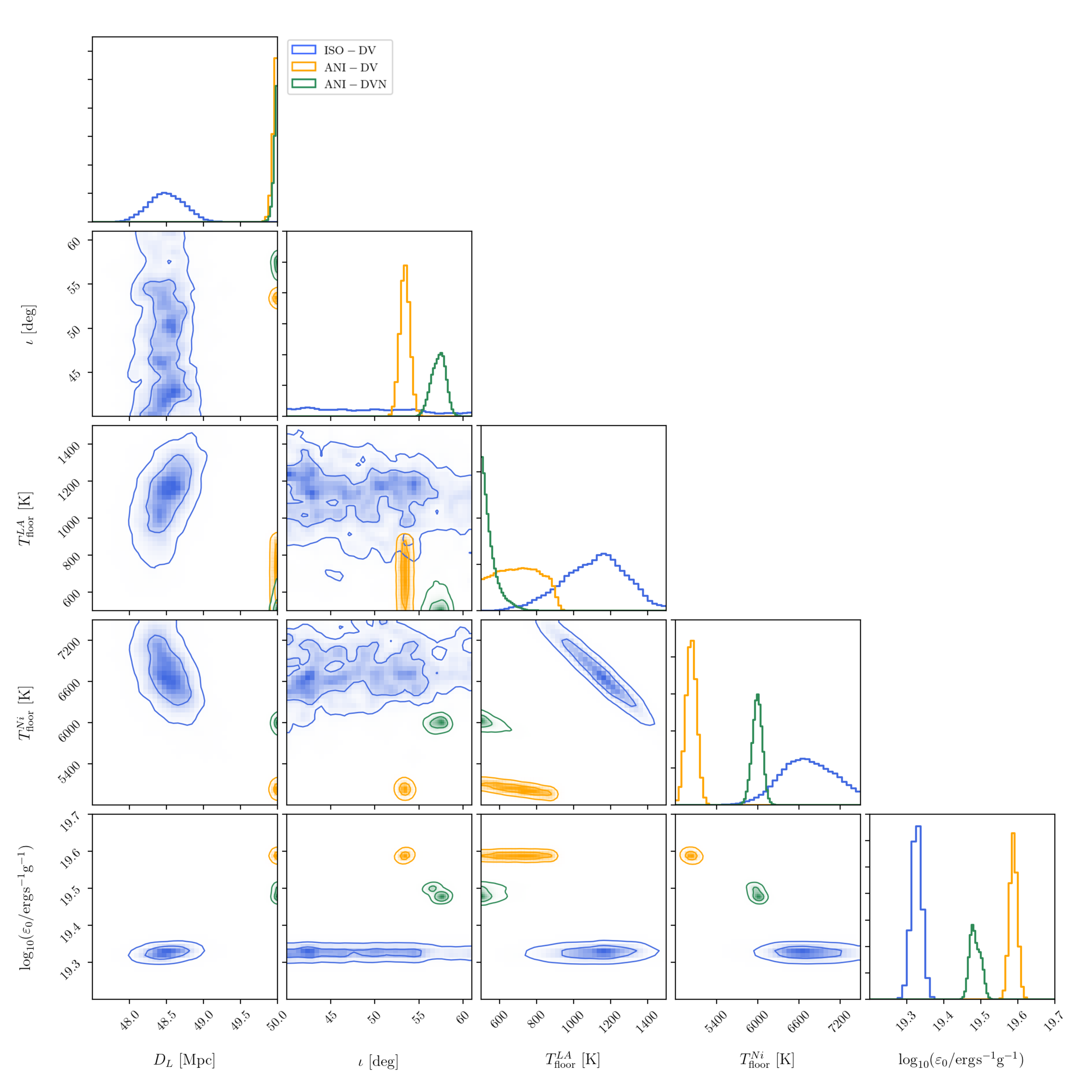}
	\caption{Marginalized posterior distribution 
		for the global intrinsic parameters and the extrinsic parameters 
		extracted from the analysis of ISO-DV, ANI-DV and ANI-DVN.
		The reported parameters are luminosity distance $D_L$,
		the viewing angle $\iota$, the floor temperatures $\Tfla$ and $\Tfni$
		and the logarithm of the heating factor $\eps_0$.
		For the ISO-DV case, the posterior distribution for the viewing angle $\iota$
		coincides with prior due to the employed geometry.}
	\label{fig:23comp_extr}
\end{figure*}

Fig.~\ref{fig:23comp_extr} shows the posterior distribution 
for the extrinsic parameters.
The temperatures $\Tfni$ and $\Tfla$ are well constrained respectively around ${\sim}5000~{\rm K}$ and 
${\sim}700~{\rm K}$.
The agreement with Ref.~\cite{Korobkin:2012uy}
	on the estimation of the heating factor $\eps_0$
	 increases with respect to the ANI-D case, 
	due to the inclusion of an additional component,
	similarly to what is discussed in Sec.~\ref{sec:iso-dv}.
The posterior for inclination angle results similar to the ANI-D case, 
according with the fact
that the viscous component, as we have defined it, does not introduce further information on the inclination.

\subsubsection{ANI-VN}
\label{sec:ani-vn}

According to Tab.~\ref{tab:evidence}, this ANI-VN is the least likely model
among all anisotropic cases. As previously mentioned, the reason for this is clear from the LCs.
The parameters of the viscous component are characterized by a slow velocity of ${\sim}6{\times}10^{-3}\,c$
and a low opacity environment, $\kappa\sim 0.5~\igscm$.
On the other hand, the neutrino-driven wind mass is overestimated compared with 
aftermath computations presented in Ref.~\cite{Perego:2017wtu}, in order to compensate the lack 
of overall ejected mass due to the absence of a dynamical component.
Moreover, the neutrino-driven wind is characterized by a realistic velocity of ${\sim}0.1\,c$,
and by a low-opaque environment, $\kappa\sim 1~\igscm$.

Regarding the extrinsic parameters, 
the ANI-VN model is the case that gives the best agreement with Ref.~\cite{Korobkin:2012uy} 
in terms of heating factor. The distance, instead, is recovered around ${\sim}20~{\rm Mpc}$,
underestimating the GW distance~\cite{TheLIGOScientific:2017qsa}.
This result could be explained by the lower amount
	of total ejected mass and by the lower heating rate compared with the other cases 
	(see Tab.~\ref{tab:intr_posteriors}): this lack generates fainter 
	kN that biases the source to appears closer to the observer in order to fit the data.
The $\Tfni$ parameter takes lower values (${\sim}3300$~K) comparing with the ANI-DV case (${\sim}5000$~K), 
since the model has to fit the data employing a polar geometry (N) instead of an equatorial ejecta (D).
The viewing angle is biased toward larger values, roughly ${\sim}50$~deg, inconsistent with GRB 
expectations~\cite{Monitor:2017mdv,Savchenko:2017ffs}.

\subsubsection{ANI-DVN}
\label{sec:ani-dvn}

This is the model that gives the largest evidence, within the provided prior bounds.
Regarding the dynamical and viscous ejecta components, 
the general features are similar to the one of the ANI-DV case.
The dynamical ejected mass is slightly overestimated comparing with NR
simulations~\cite{Perego:2019adq,Nedora:2019jhl,Endrizzi:2019trv,Nedora:2020pak,Bernuzzi:2020txg}
of a factor ${\sim}2$.
The dynamical component is described by a low opacity
environment for high-latitudes ($\khigh\sim 0.1~\igscm$) and 
high opacity for low-latitudes ($\klow\sim 11~\igscm$),
in agreement with NR simulations~\cite{Perego:2019adq,Nedora:2019jhl,Endrizzi:2019trv,Nedora:2020pak,Bernuzzi:2020txg}.
These results approximately agree also with other observational estimations~\citep[e.g.,][]{Villar:2017wcc,  Cowperthwaite:2017dyu,Abbott:2017wuw,Coughlin:2018miv}
Furthermore, the `D' component results into
the fasted ejected shell, validating the 
interpretation that this contribution is generated at dynamic time-scales.
On the other hand,
the viscous ejecta is characterized by an average opacity ${\sim}3~\igscm$ and
by low velocity ${\sim}3{\times}10^{-3}\,c$, an order of magnitude smaller then the one
of the dynamical ejecta.
These results agree with the studies presented in 
Ref.~\cite{Radice:2018ghv} and they contribute to
  the LCs in the optical band.

\begin{figure}
	\centering 
	\includegraphics[width=0.49\textwidth]{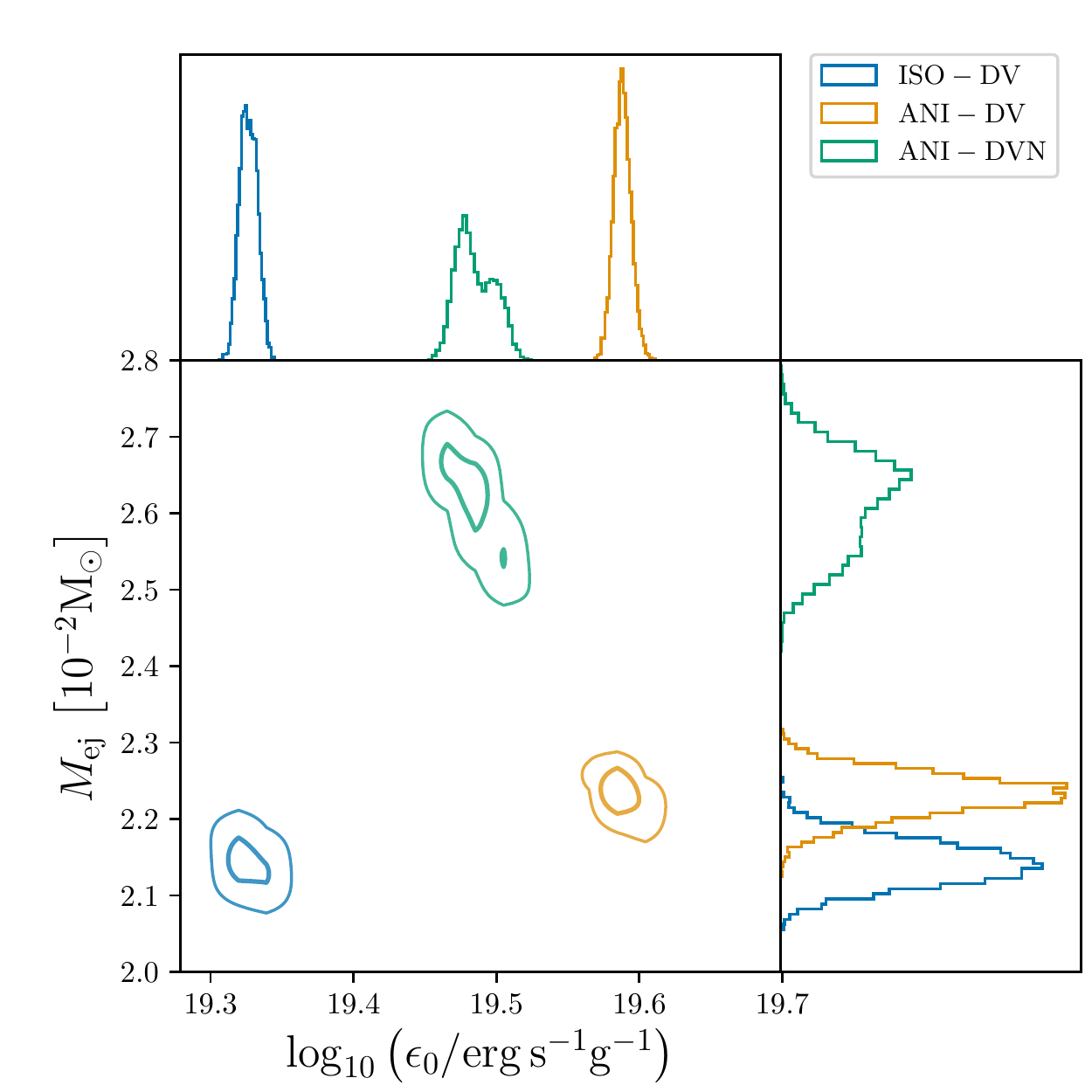}
	\caption{Marginalized posterior distribution 
					of heating parameter $\epsilon_0$
					and total ejected mass 
					$\Metot$ for three selected cases:
					ISO-DV (blue), ANI-DV (yellow) and ANI-DVN (green).
					The heating parameter $\eps_0$ is plotted using the logarithm to base 10
					in order to evince the recovered orders of magnitude.
					The total mass $\Metot$ is computed extending the sum to all the involved components.}
	\label{fig:Me}
\end{figure}

Regarding the neutrino-driven wind,
the posterior distribution for its ejected mass $\Me^{\rm (N)}$ shows a bimodality 
and this degeneracy correlates with the heating rate parameter $\eps_0$.
This behavior can be seen in Fig.~\eqref{fig:Me},
that shows the marginalized posterior distribution for $\eps_0$ and 
for the total ejected mass $\Metot$, defined as
\be
\label{eq:mej_tot}
\Metot = \sum_{k={\rm D,N,V}} \Me^{(k)}\,,
\ee
where the index $k$ runs over all the involved components.
The marginalized posterior distribution for $\Me^{\rm (N)}$ has its dominat peak 
in proximity of $2.5{\times}10^{-3}~\Mo$, 
while the secondary mode is located slightly below $2{\times}10^{-3}~\Mo$.
Despite the bimodality, the recovered values of $\Me^{\rm (N)}$ are smaller compared with 
the same parameter extracted from the ANI-VN analysis.
These results are largely consistent with aftermath computations~\cite{Perego:2014fma} and with  
theoretical expectations~\cite{Perego:2017wtu}, as it is for the
recovered velocity and opacity parameters.

Furthermore, also for the ANI-DVN case, the viewing angle is biased toward larger values, roughly ${\sim}60$~deg.
The same trend is shown by the anisotropic three-component model employed in Ref.~\cite{Villar:2017wcc}.
The posterior distribution for the $\Tfni$ parameter peaks around ${\sim}6000~{\rm K}$, while, 
the temperature $\Tfla$ is constrained around the lower bound, $500~{\rm K}$.

\section{EOS Inference}
\label{sec:eos}

The combination of gravitational and electromagnetic signals coming from the same compact binary merger
allows the possibility to constrain more tightly the intrinsic properties of the system and
the nuclear EOS, in the context of both BNS (e.g., \citet{Radice:2018ozg,Radice:2017lry})
and black hole-NS mergers (e.g., \citet{Barbieri:2019sjc}).
In this section, we apply the information coming from NR fitting formulae~\cite{Nedora:2020pak,Nedora:2020qtd}
to the posterior distribution of the 
preferred kN model (ANI-DVN), in order to 
infer the mass ratio and the reduced tidal parameter of the BNS source.
Subsequently, we combine the kN and GW results to derive constraints on the radius $\rof$
of an irrotational NS of 1.4~$\Mo$.

\subsection{Mass ratio and reduced tidal parameter}

A BNS is characterized by the masses of the two objects, 
$m_1$ and $m_2$, 
and by the tidal quadrupolar polarizability coefficients,
\be
\label{eq:lambdacomponent}
\Lambda_i = \frac{2}{3}\,k_{2,i} C_{i}^{-5}\,,
\ee
where $k_{2,i}$ is the quadrupolar Love number, 
$C_i=G m_i/(R_i c^2)$ 
the compactness of star, $G$ the gravitational constant, $R_i$ the radius of the star and $i=1,2$.
Furthermore, we introduce the mass ratio $q = m_1/m_2\ge 1$ and the 
reduced tidal parameter $\lt$ as:
\be
\label{eq:lambdat}
\lt = \frac{16}{13} \, \frac{(q+12)q^4 \Lambda_1 +(1+12q)\Lambda_2 }{(1+q)^5}\,.
\ee
The NR fits presented in Ref.~\cite{Nedora:2020qtd} use simulations
targeted to GW170817
\cite{Perego:2019adq,Endrizzi:2019trv,Nedora:2019jhl,Bernuzzi:2020txg,Nedora:2020pak}
and give the mass $\Me^{\rm (D)}$ and velocity $\vrms^{\rm (D)}$ of the dynamical ejecta 
as functions of the BNS parameters $(q,\lt)$.
In order to recover the posterior distribution of the latter, we adopt a resampling method, similar 
to the procedure presented in Ref.~\cite{Coughlin:2017ydf,Coughlin:2018miv}:
a sample $(q, \lt)$ is extracted from 
the prior distribution
\footnote{The prior distribution is taken uniformly distributed in the tidal parameters
		$\lt$; while, regarding the mass ratio $q$, we employ
		a prior distribution uniform in the mass components,
		that corresponds to 
					a probability density proportional to $[(1+q)/q^3]^{2/5}$,
					analogously to GW analyses~\cite{TheLIGOScientific:2017qsa,Gamba:2020ljo}.},
exploiting the ranges $q\in[1,2]$ and $\lt \in[0,5000]$.
Subsequently, the tuple $(q, \lt)$ is
mapped into the dynamical ejecta parameters $(\Me^{\rm (D)},\vrms^{\rm (D)})$
using the NR formulae presented in Ref.~\cite{Nedora:2020qtd}.
The likelihood is estimated in the dynamical ejecta parameter space
using a kernel density estimation 
of the marginalized posterior distribution 
recovered from the preferred model (ANI-DVN).
Furthermore, since NR relations have
non-negligible uncertainties,
we introduce calibration parameters $\alpha_1, \alpha_2$,
such that 
\be
\label{eq:calib}
\begin{split}
\log_{10}\Me^{\rm (D)}  &= (1+\alpha_1)\cdot\log_{10}\Me^{\rm (D)} _{\rm fit}(q,\lt)\,,\\
\vrms^{\rm (D)} &= (1+\alpha_2)\cdot\vrms^{\rm (D)} _{\rm fit}(q,\lt)\,.\\
\end{split}
\ee
The calibrations parameters $\alpha_{1,2}$ are sampled along the other parameters
using a normally distributed prior with vanishing means and standard deviations 
prescribed by the relative uncertainties of NR fits equal to
$0.2$ for both. 
The resampled posterior distribution is 
marginalized over the calibration parameters.
The BNS parameter space is explored using a Metropolis-Hasting technique.
Note that a correct characterization of the fit uncertainty is crucial,
since this contribution is the largest source of error in the inference of $(q,\lt)$.

\begin{figure}
	\centering 
	\includegraphics[width=0.49\textwidth]{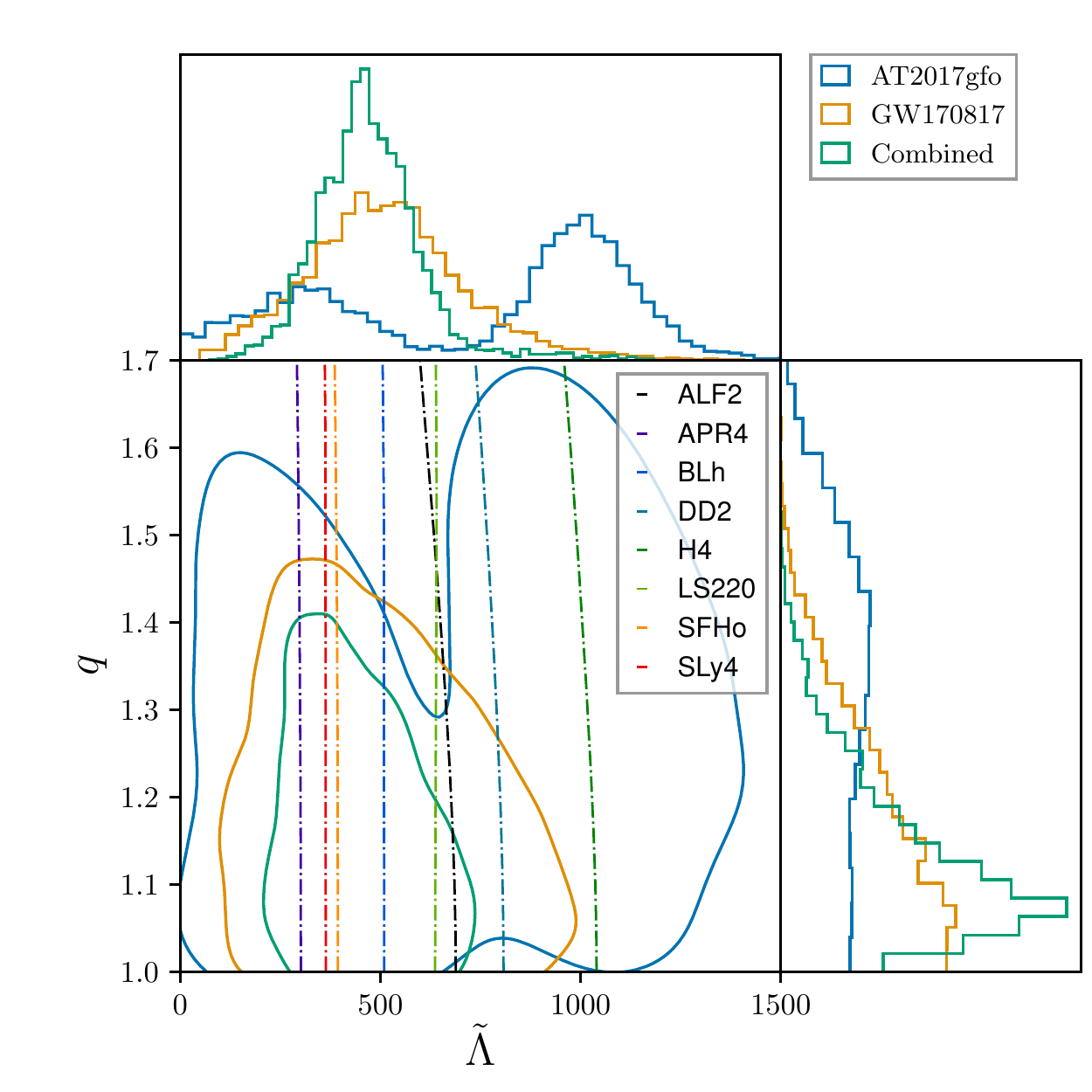}
	\caption{Posterior distribution in the $(\lt,q)$ plane.
					The blue solid lines refer to the resampled values 
					extracted from the kN analysis (ANI-DVN).
					The orange solid lines refer to the GW results,
					 where the samples have been reweighted 
					over a flat prior in $\lt$.
					The green solid lines are the combined inference.
					The contours represent the 90\% credible regions
					The plot shows the expectations of 
					some representative EOS.
	}
	\label{fig:ltq}
\end{figure}

The posterior distribution in the $(q,\lt)$ plane 
as obtained from the dynamical ejecta properties fitted to AT2017gfo data
is shown in Fig.~\ref{fig:ltq}.
The measurement of the tidal parameter leads to $\lt=900^{+310}_{-780}$,
with a bimodality in the marginalized posterior distribution, 
due to the quadratic nature of the employed NR formulae, 
with modes $\lt\sim370$ and $\lt\sim1000$. 
The mass ratio is constrained to be lower than $1.54$ at the
90\% confidence level. The uncertainties of these estimations 
are larger 
than those of the GW analyses~\cite{TheLIGOScientific:2017qsa,Abbott:2018wiz,Gamba:2020ljo}
and the principal source of error is the uncertainty of the NR fit formulae. 

Fig.~\ref{fig:ltq} shows also the results coming from the 
GW170817 analysis extracted from Ref.~\cite{Gamba:2020ljo}.
For this analysis, the data correspond to the LIGO-Virgo strains~\cite{TheLIGOScientific:2017qsa,Abbott:2018wiz,LIGOScientific:2018mvr} centered around GPS time 1187008882 with sampling rate of 4096~Hz and duration of 128~s.
The parameter estimation has been performed with the nested sampling provided by the
{\tt pbilby} pipeline~\cite{Ashton:2018jfp, Smith:2019ucc} 
employing the effective-one-body waveform approximant {\tt TEOBResumSPA}~\cite{Nagar:2018zoe,Gamba:2020ljo}
and analyzing the frequency range from 23~Hz to to 1024~Hz
~\footnote{This choice minimizes waveform systematics~\cite{Gamba:2020wgg}. On the other hand, it
  implies slightly larger statistical uncertainties on the reduced
  tidal parameters. Hence, our results are more conservative than
  previous multimessenger analyses in the treatment of uncertainties
  of GW data.}. 
Furthermore, the GW posterior samples 
have been reweighted with a rejection sampling
to the prior distributions employed in the kN study,
in order to use the same prior information 
for both analyzes
\footnote{The prior distribution for the tidal parameters
	employed in Ref.~\cite{Gamba:2020ljo} is uniform in the 
	tidal components $\Lambda_{1,2}$; 
	while, in our study, we used a
	uniform prior in $\tilde\Lambda$.}. 

Under the assumption that GW170817 and AT2017gfo are 
generated by the same physical event,
the $(q,\lt)$ posterior distributions coming from the 
two independent analyses can be combined, in order to 
constrain the estimation of the inferred quantities. 
The joint probability distribution is 
computed as the product of the single terms,
\be
p\big(q,\lt\big|d_{\rm kn},d_{\rm gw}\big) =
 p\big(q,\lt\big|d_{\rm kn}\big) \cdot p\big(q,\lt\big|d_{\rm gw}\big)\,,
\ee
and the samples are extracted with a rejection sampling.
The combined inference, shown in Fig.~\ref{fig:ltq}, 
leads to a constraint on the mass ratio of ${\lesssim}1.27$
and on the tidal parameter $\lt = 460^{+210}_{-190}$,
at the  90\% confidence levels.
Imposing these bounds, stiff
nuclear EOS, such as DD2, are disfavored.

\subsection{Neutron-star radius}

\begin{table}
	\centering    
	\caption{Estimated values of mass ratio $q$,
					reduced tidal parameter $\lt$ and NS radius $\rof$
					measured from the analyses of AT2017gfo and GW170817.
					The $\rof$ are estimated using the relation proposed in 
					Ref.~\cite{De:2018uhw,Radice:2018ozg} and employing the 
					chirp mass posterior distribution coming from the GW analysis~\cite{Gamba:2020ljo}.}
		\begin{tabular}{c|ccccc}        
			\hline\hline
			Data & $q$ & $\lt$ & $\rof$ \\
			 &  &  & [km] \\
			\hline\hline
			AT2017gfo & ${\le}1.54$ & $900^{+310}_{-780}$ &$13.46^{+0.93}_{-3.82}$\\
			GW170817 & ${\le}1.33$ & $510^{+350}_{-320}$&$12.33^{+1.22}_{-1.85}$\\
			Combined & ${\le}1.27$& $460^{+210}_{-190}$&$12.16^{+0.89}_{-1.11}$\\
			\hline\hline
		\end{tabular}
	        \label{tab:r14}\\
\end{table}

Using the universal relation presented in Ref.~\cite{De:2018uhw,Radice:2018ozg},
it is possible to impose a constraint on the radius $\rof$ of a 
NS of $1.4~\Mo$.
We employ the marginalized posterior distribution
for the (source-frame) chirp mass $\mathcal{M} = (m_1 m_2)^{3/5}/(m_1+m_2)^{1/5}$
coming from the GW170817 measurement~\cite{Gamba:2020ljo} and the  
posterior on the tidal parameter $\lt$ obtained with the joint analyses 
AT2017gfo+GW170817.
We adopt a resampling technique 
to account for the uncertainties in the universal relation, 
introducing a Gaussian calibration
coefficient with variance prescribed by Ref.~\cite{De:2018uhw,Radice:2018ozg}.
We estimate $\rof=12.16^{+0.89}_{-1.11}~{\rm km}$.
The presented measurement
agrees with the results coming from literature~\cite{Annala:2017llu,De:2018uhw,Radice:2018ozg,
	Coughlin:2018fis,Abbott:2018exr,
	Raaijmakers:2019dks,Capano:2019eae,Essick:2020flb,
	Dietrich:2020efo} 
and its overall error at $1\sigma$ level
corresponds roughly to $500~{\rm m}$.

\begin{figure}
	\centering 
	\includegraphics[width=0.49\textwidth]{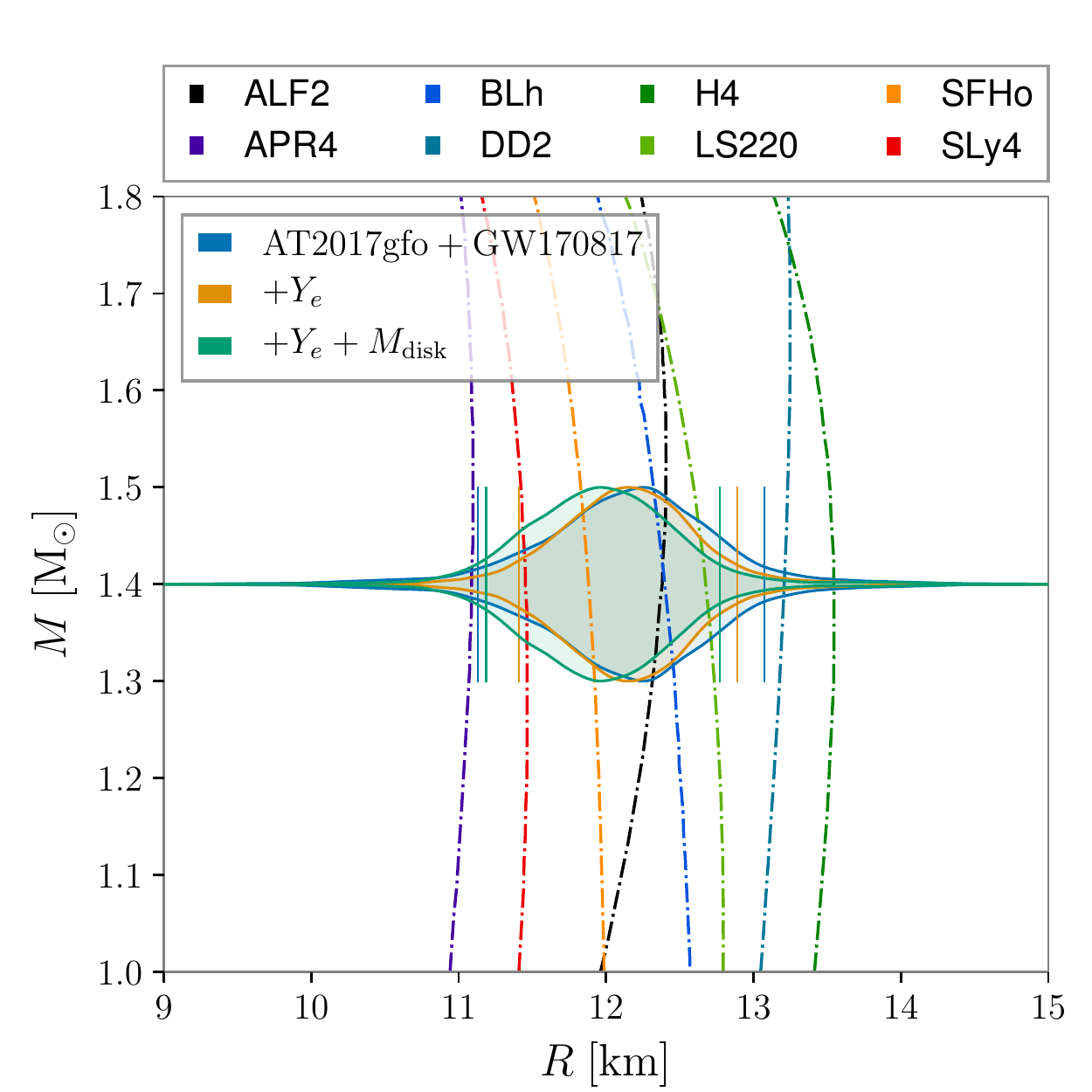}
	\caption{Posterior distribution of the radius $\rof$ 
	  estimated with the joined inference of AT2017gfo and GW170817
	  plotted on top of the mass-radius
	  relations coming from a sample of nuclear EOS
	  (dashed lines).
	  The blue solid line is computed using the mass and velocity 
	  information of the dynamical component,
	  the orange solid curve takes into account also the contribution of the electron
	  fraction and the green solid line is the result with the additional inclusion of the 
	  disk mass information.
	}
	\label{fig:r14}
\end{figure}

In Fig.~\ref{fig:r14}, the $\rof$ estimation is compared with the mass-radius curves 
from a sample of nuclear EOS. 
Our bounds impose observational constraints on the 
nuclear EOS, excluding both very stiff EOS, such as DD2, 
BHB$\Lambda\phi$ and MS1b, and very soft 
equations, such as 2B.

\subsection{Incorporating information from electron fraction and disk mass}

We conduct two further analyses, 
in order to show that the contribution of additional NR information
can improve the previous estimation. 
In the first case, we take into account the 
contribution of the electron fraction; 
while, in the second, we include the information
on the disk mass. 
These studies are discussed in the following 
paragraphs and they are intended to represent proofs-of-principle
analyses, since 
they involve extra assumptions on the 
ejecta parameters and their relation with the EOS properties.
A more accurate mapping between these quantities will be discussed in a 
further study.

\subsubsection{Electron fraction}
\label{sec:yefit}

\begin{figure}
	\centering 
	\includegraphics[width=0.49\textwidth]{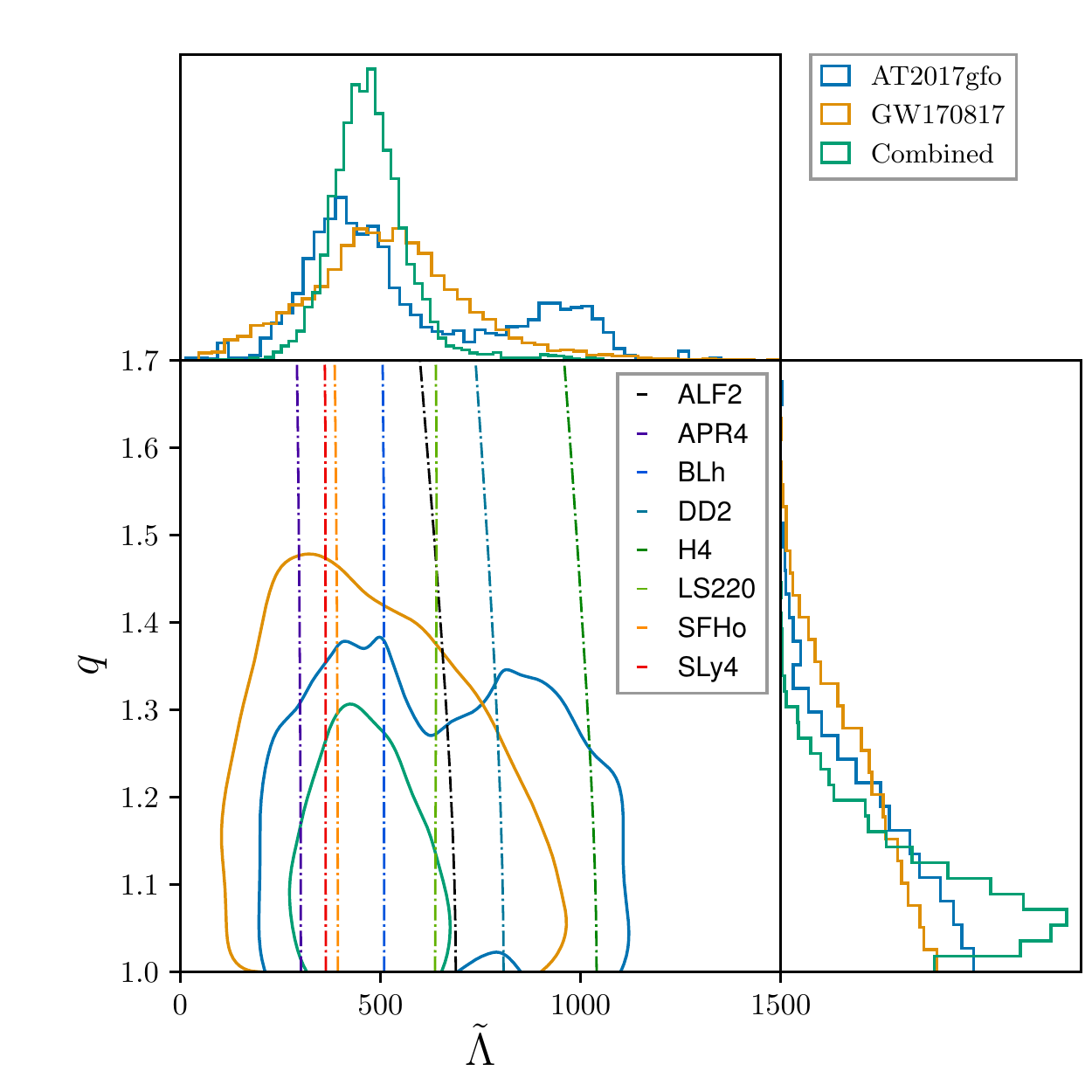}
	\caption{Posterior distribution in the $(\lt,q)$ plane,
		analogously to Fig.~\ref{fig:ltq}, 
		including the contribution of the electron fraction $Y_e$.
	}
	\label{fig:ltq+ye}
\end{figure}

From NR simulations, it is possible to estimate the average
electron fraction, $Y_e$, of the dynamical ejecta~\cite{Nedora:2020pak,Nedora:2020qtd}.
This quantity is the ratio of the net number of electrons to 
the numer of baryons and it is strictly related with the opacity of the 
shell~\cite{Lippuner:2015gwa,Miller:2019dpt,Perego:2019adq}, 
since it mostly determines the nucleosynthesis yields in low entropy, neutron-rich
matter. 
We compute the average opacity $\bar\kappa$ of a shell
as the integral of the opacity over the polar angle weighted on the mass 
distribution,
\be
\bar\kappa = \frac{1}{\Me}\int_0^{\pi}
\varrho(\theta)\,\kappa(\theta)\,\sin\theta\,{\rm d}\theta\,.
\ee
Imposing the assumptions on the profiles of the dynamical ejecta, we get
\be
\bar\kappa^{\rm (D)} =\left( \frac{1}{2}+\frac{1}{\pi} \right)\klow^{\rm (D)} +\left( \frac{1}{2}-\frac{1}{\pi} \right)\khigh^{\rm (D)}\,.
\ee
Thanks to this definition, it is possible to map the opacity $\bar\kappa$
into the electron fraction $Y_e$, using the relation presented 
in Ref.~\cite{Tanaka:2019iqp}. Subsequently, 
 the $Y_e$ can be related with the BNS parameters $(q,\lt)$, using
 NR fit formulae~\cite{Nedora:2020qtd}.
 We introduce an additional calibration parameter $\alpha_3$,
 such that
 \be
 Y_e = (1+ \alpha_3)\cdot {Y_e}^{\rm fit}(q,\lt)\,,
 \ee
 with a Gaussian prior with mean zero and standard deviation of $0.2$.
In this way it is possible to take into account also the contribution
 of the opacity posterior distribution, introducing additional constraints 
on the inference of the NS matter.

The results are shown in Fig.~\ref{fig:ltq+ye}.
This further contribution has a strong effect on the mass
ratio, constraining it to be ${\lesssim}1.26$. This effect
is motivated by the fact that high-mass-ratio BNS mergers are expected
to have $Y_e\lesssim 0.1$~\cite{Bernuzzi:2020txg,Nedora:2020pak}.
The recovered electron fraction correspond to $Y_e=0.20^{+0.03}_{-0.05}$.
Regarding the tidal parameter, the $Y_e$ information 
affects the importance of the modes, improving the agreement with 
GW estimations~\cite{TheLIGOScientific:2017qsa,Abbott:2018wiz,Gamba:2020ljo}, and it reduces the support of the posterior distribution,
leading to an estimation of $\lt=480^{+550}_{-220}$.
Combining kN and GW posterior distribution, we estimate 
an upper bound on the mass ratio of $1.20$ and a tidal parameter
$\lt=465^{+175}_{-130}$, that corresponds to $\rof=12.14^{+0.75}_{-0.73}~{\rm km}$,
at the 90\% confidence level.

\subsubsection{Disk mass}
\label{sec:diskfit}

\begin{figure}
	\centering 
	\includegraphics[width=0.49\textwidth]{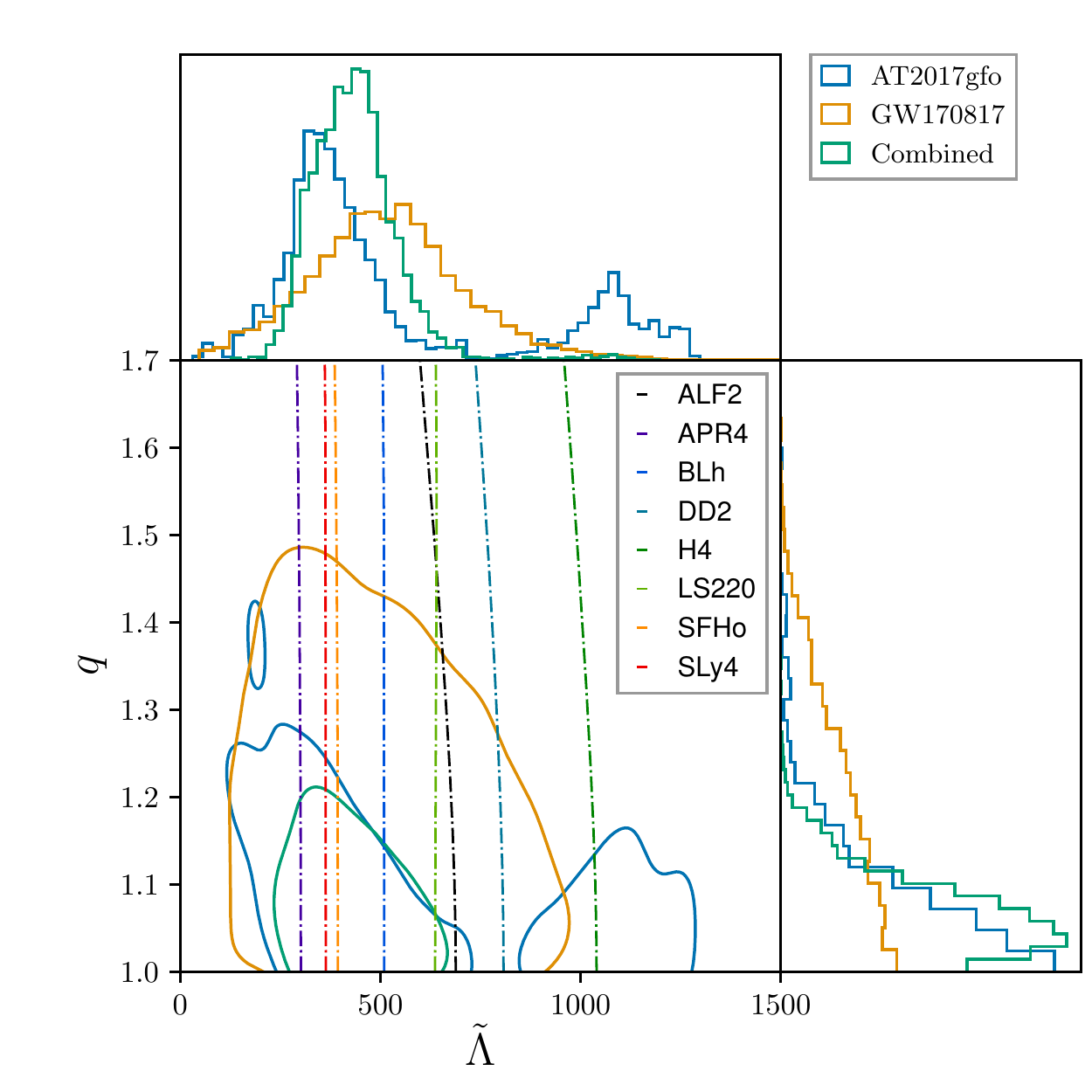}
	\caption{Posterior distribution in the $(\lt,q)$ plane,
		analogously to Fig.~\ref{fig:ltq}, 
		including the contributions of electron 
		fraction $Y_e$ and disk mass $M_{\rm disk}$.
	}
	\label{fig:ltq+disk+ye}
\end{figure}

The employed kN model contains information also on the 
baryonic wind ejecta.
These components are expected to be generated by the disk
that surrounds the remnant~\cite{Kasen:2014toa,Metzger:2014ila,Just:2015fda}, if present.
The disk mass can be estimated from NR simulations as function of the BNS parameters $(q,\lt)$, albeit with large uncertainties ~\cite{Radice:2017lry,Radice:2018pdn,Nedora:2020qtd}.
We map a fraction $\xi$ 
of the disk mass $M_{\rm disk}$ into the mass of the 
baryonic wind components,
\be
\Me^{\rm (V)} + \Me^{\rm (N)} = \xi \cdot M_{\rm disk}\,.
\ee
The mass fraction $\xi$ is sampled along the other parameters with 
a uniform prior in the range $[0.1,0.5]$.
We include the disk mass information together with the 
electron fraction contribution, previously discussed.

The results are shown in Fig.~\ref{fig:ltq+disk+ye}.
The disk mass contribution slightly reinforces the 
constraint on the mass ratio posterior, 
giving the 90\% confidence level for $q=1.18$.
The distribution of the tidal parameter $\lt$ is sparser
with respect to the case discussed in Sec.~\ref{sec:yefit}, due to the 
correlations induced by the $M_{\rm disk}$ formula.
The electron fraction results $Y_e={0.20}^{+0.04}_{-0.08}$;
while, the mass fraction
corresponds to $\xi={0.14}^{+0.27}_{-0.04}$.
The joined inference with the GW posterior leads to a 
mass ratio $\lesssim 1.13$ and a tidal parameter of $\lt=430^{+180}_{-140}$,
at the 90\% confidence. This result can be translated in a radius of 
$\rof=11.99^{+0.82}_{-0.85}~{\rm km}$.


\section{Conclusion}
\label{sec:conclusions}

In this paper, we have performed informative model selection on kN observations within 
a Bayesian framework applied to the case of AT2017gfo, the kN associated with
the BNS merger GW170817. We have then combined the posteriors obtained from
the kN observation with the ones extracted from the GW signal and with NR-based
fitting formulae on the ejecta and remnant properties to set tight constraints
on the NS radius and EOS. 

From the analysis of AT2017gfo,
the anisotropic description of the ejecta components
is strongly preferred with respect to isotropic profiles,
with a logarithmic Bayes' factor of the order of ${\sim}10^4$.
Moreover, the favored model is the three-component kN 
constituted by a fast dynamical ejecta (comprising both a
red-equatorial and a blue-polar portion), a slow isotropic shell and a
polar wind.  
For the best model, the dynamical ejected mass overestimates of a factor two
the theoretical expectation coming from NR
simulations~\cite{Perego:2019adq,Nedora:2019jhl,Endrizzi:2019trv,Nedora:2020pak,Bernuzzi:2020txg}.
These biases can be explained by considering the effect of the spiral-wave wind~\cite{Nedora:2019jhl} and  
taking into account the correlations between the extrinsic parameters.
The recovered velocity of the dynamical component 
agrees with NR simulations~\cite{Perego:2019adq,Nedora:2019jhl,Endrizzi:2019trv,Nedora:2020pak,Bernuzzi:2020txg}, 
reinforcing the interpretation of this ejecta component.
The intrinsic properties of the dynamical ejecta component are in
agreement with previous results~\cite{Villar:2017wcc,Coughlin:2018fis}.
Regarding the secular winds, 
the neutrino-driven 
mass and velocity are compatible with the calculations of 
Ref.~\cite{Perego:2014fma,Perego:2017wtu}. 
The viscous component is the slowest contribution and is broadly compatible
with the estimates of Ref.~\cite{Radice:2018ghv}, that are inferred
from NR and other disc simulations.
The viewing angle resulting from the preferred kN model is larger than
the one deduced from independent analysis \cite{Monitor:2017mdv,Savchenko:2017ffs,Ghirlanda:2018uyx}, and also different from the one obtained
by previous application of the same kN model \cite{Perego:2017wtu}. In the latter case, and differently
from the present analysis, the profile of the viscous ejecta was assumed to be mostly distributed 
across the equatorial angle. This discrepancy confirms the non-trivial dependence of the light curves
from the ejecta geometry and distributions.

Under a modeling perspective,
current kN description contains large theoretical uncertainties,
such as thermalization effects, heating rates and energy-dependent
photon opacities, e.g. \cite{Zhu:2020eyk}. These effects propagate
into systematic biases in the global parameters of the model,
as shown in the posterior distributions for luminosity distance $D_L$
and heating rate parameter $\eps_0$.
Hence, the development and the improvements of kN
templates is an urgent task in order to conduct reliable and robust analyses
in the future.
	
	\begin{figure}
		\centering 
		\includegraphics[width=0.49\textwidth]{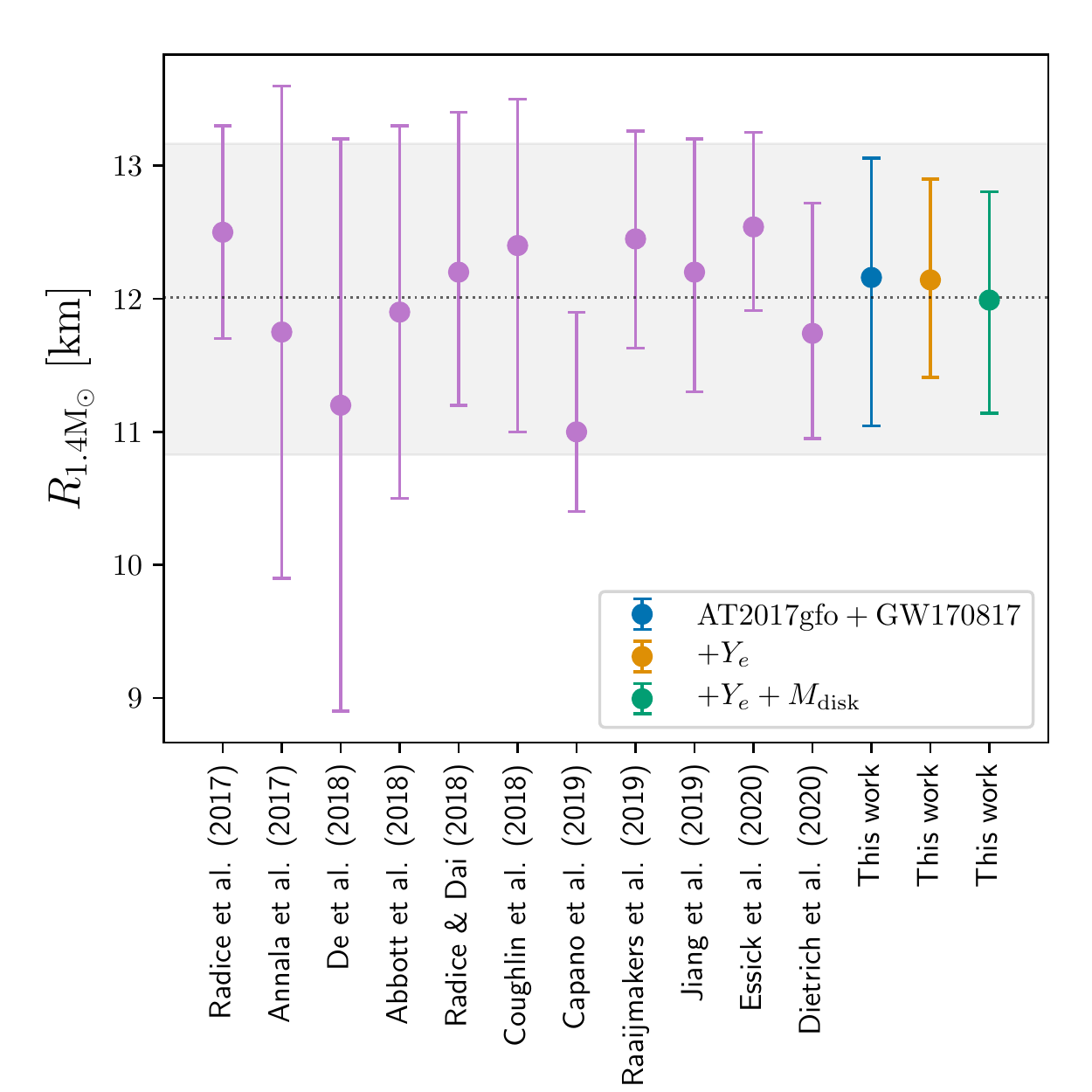}
		\caption{Summary plot of the current estimations of $\rof$.
			The reported values are the means and the 90\% credible regions
			extracted from Refs.~\cite{Annala:2017llu,Radice:2017lry,
				De:2018uhw,Radice:2018ozg,
				Coughlin:2018fis,Abbott:2018exr,
				Raaijmakers:2019dks,Capano:2019eae,
				Jiang:2019rcw,Essick:2020flb,Dietrich:2020efo}.
			The dashed line and the shadowed area are respectively the average over all the current
			estimations and the respective 90\% credible region,
			corresponding to $\rof={12.0}^{+1.2}_{-1.2}$~km.}
		\label{fig:r14:history}
	\end{figure}

We use of the preferred kN model to constrain the properties of the
progenitor BNS and the EOS of dense, cold matter. 
Combining the kN measurement with the information coming from NR simulations,
the ejecta properties are mapped in terms of 
mass ratio and reduced tidal deformability of the binary progenitor.
Subsequently, this information is
combined with the measurements of the GW data. 
The joint kN+GW analysis constrains 
the reduced tidal parameter to $\lt = 460^{+210}_{-190}$ and  the mass ratio 
of the BNS system to be lower than $1.27$, at the 90\% credible level. 
Furthermore, the joint analysis predicts a radius for a NS of $1.4~\Mo$ approximately 
of $\rof\approx 12.2~{\rm km}$ with an uncertainty of ${\sim}500~{\rm m}$ at one-$\sigma$ level. 
The $\rof$ estimation can be further improved including
additional physical information extracted from the kN model in the inferred model, 
such as the electron fraction of the dynamical ejecta
and the mass of the disk around the merger remnant. 
Figure~\ref{fig:r14:history} summarizes ours and the current estimations of $\rof$ 
extracted from literature~\cite{Annala:2017llu,Radice:2017lry,De:2018uhw,Radice:2018ozg,
 		Coughlin:2018fis,Abbott:2018exr,
 		Raaijmakers:2019dks,Capano:2019eae,Jiang:2019rcw,Essick:2020flb,
 		Dietrich:2020efo}.

In addition to the kN modeling uncertainties discussed above,
another source of error of our estimates is 
the accuracy of the NR formulae.
The relations employed here used exclusively targeted data and
simulations with state-of-art treatment of microphysical EOS and neutrino treatment
\cite{Perego:2019adq,Nedora:2019jhl,Endrizzi:2019trv,Nedora:2020pak,Bernuzzi:2020txg}. However,
the simulation sample is limited to about hundrends of simulations,
with fitting errors that could be reduced by considering data at even
higher grid resolutions~\cite{Nedora:2020qtd}. 
For example, assuming all the fit formulae to be exact
(i.e. removing all calibration terms), 
it will be possible to infer the $\lt$ parameter from a kN observation
with an accuracy of the order of 10,
that corresponds to a constraint on the radius $\rof$ of roughly $100~{\rm m}$.


\section*{Acknowledgements}
  M.B. and S.B. acknowledges support by the European Union's
  H2020 under ERC Starting Grant, grant 
  agreement no. BinGraSp-714626. 
  D.R. acknowledges support from the U.S. Department of Energy, Office
  of Science, Division of Nuclear Physics under Award Number(s)
  DE-SC0021177 and from the National Science Foundation under Grant No.
  PHY-2011725.
  The computational experiments were performed on the ARA cluster at Friedrich
  Schiller University Jena supported in part by DFG grants INST
  275/334-1 FUGG and INST 275/363-1 FUGG, and ERC Starting Grant, grant 
  agreement no. BinGraSp-714626.
  Data postprocessing was performed on the Virgo “Tullio”
  server at Torino supported by INFN.
  This research has made use of data, software and/or web tools obtained from the Gravitational Wave Open Science Center (\href{https://www.gw-openscience.org/}{\tt https://www.gw-openscience.org}), a service of LIGO Laboratory, the LIGO Scientific Collaboration and the Virgo Collaboration. LIGO Laboratory and Advanced LIGO are funded by the United States National Science Foundation (NSF) as well as the Science and Technology Facilities Council (STFC) of the United Kingdom, the Max-Planck-Society (MPS), and the State of Niedersachsen/Germany for support of the construction of Advanced LIGO and construction and operation of the GEO600 detector. Additional support for Advanced LIGO was provided by the Australian Research Council. Virgo is funded, through the European Gravitational Observatory (EGO), by the French Centre National de Recherche Scientifique (CNRS), the Italian Istituto Nazionale della Fisica Nucleare (INFN) and the Dutch Nikhef, with contributions by institutions from Belgium, Germany, Greece, Hungary, Ireland, Japan, Monaco, Poland, Portugal, Spain.


\section*{Data availability}
The observational data underlying this article were provided by \cite{Villar:2017wcc} under license. 
The posterior samples presented in this work will be shared on request to the corresponding author.


\bibliography{references}

\end{document}